\documentclass[twocolumn,twocolappendix]{aastex701}

\usepackage{enumitem}
\usepackage{xcolor}

\usepackage{booktabs}

\usepackage{siunitx}
\begin{document}

\title{A MUSE View of the Optical Torus within the Supernova Remnant 1E~0102.2--7219}

\author[orcid=0000-0001-7068-9702,sname='Suherli']{Janette Suherli}
\affiliation{Department of Physics and Astronomy, The University of Manitoba, Winnipeg, MB~R3T~2N2, Canada}
\email[show]{suherlij@myumanitoba.ca}  

\author[orcid=0000-0001-6189-7665]{Samar Safi-Harb}
\affiliation{Department of Physics and Astronomy, The University of Manitoba, Winnipeg, MB~R3T~2N2, Canada}
\email{Samar.Safi-Harb@umanitoba.ca}

\author[orcid=0000-0002-5044-2988]{Ivo R. Seitenzahl} 
\affiliation{Research School of Astronomy and Astrophysics, Australian National University, Canberra, ACT 2611, Australia}
\email{ivo.seitenzahl@anu.edu.au}

\author[orcid=0000-0002-9665-2788]{Frédéric P. A. Vogt}
\affiliation{Federal Office of Meteorology and Climatology -- MeteoSwiss, Payerne CH-1530, Switzerland}
\email{frederic.vogt@alumni.anu.edu.au}

\author[orcid=0000-0002-9886-0839]{Parviz Ghavamian}
\affiliation{Department of Physics, Astronomy and Geosciences, Towson University, Towson, MD 21252, USA}
\email{pghavamian@towson.edu}

\author[orcid=0000-0002-6620-7421]{Ralph Sutherland}
\affiliation{Research School of Astronomy and Astrophysics, Australian National University, Canberra, ACT 2611, Australia}
\email{Ralph.Sutherland@anu.edu.au}

\author[orcid=0000-0003-1449-7284]{Chuan-Jui Li}
\affiliation{Graduate Institute of Applied Physics, National Chengchi University, Taipei 116026, Taiwan}
\email{cjli@g.nccu.edu.tw}

\author[orcid=0000-0002-4794-6835]{Ashley J. Ruiter}
\affiliation{Mathematical Sciences Institute, The Australian National University, Acton, ACT 2601, Australia}
\email{ashley.ruiter@anu.edu.au}

\author[orcid=0000-0002-4231-8717]{Gilles Ferrand}
\affiliation{Department of Physics and Astronomy, The University of Manitoba, Winnipeg, MB~R3T~2N2, Canada}
\affiliation{RIKEN Center for Interdisciplinary Theoretical and Mathematical Sciences (iTHEMS), Wak\={o}, Saitama 351-0198, Japan}
\email{Gilles.Ferrand@umanitoba.ca}

%\collaboration{all}{The Terra Mater collaboration}

%% Use the \collaboration command to identify collaborations. This command
%% takes an optional argument that is either a number or the word "all"
%% which tells the compiler how many of the authors above the command to
%% show. For example "\collaboration[all]{(DELVE Collaboration)}" wil include
%% all the authors above this command.
%%
%% Mark off the abstract in the ``abstract'' environment. 
\begin{abstract}
We present new MUSE Narrow Field Mode with Adaptive Optics observations of the optical torus surrounding a Central Compact Object (CCO) candidate within the oxygen-rich supernova remnant 1E~0102.2--7219 (E0102) located in the Small Magellanic Cloud. These data provide nearly an order-of-magnitude improvement in spatial resolution over previous MUSE Wide Field Mode observations. The improved spatial resolution resolved the previously identified torus into a cavity-like structure with a sharply defined inner edge and diffuse, outer filamentary substructure. The emission shows continuous velocity connectivity, broad intrinsic line widths, and co-spatial contributions from neutral and partially ionized species, including \ion{O}{1}, \ion{Ne}{1}, [\ion{O}{1}], [\ion{O}{2}], and [\ion{O}{3}]. Spatially resolved line-ratio maps indicate that the emission arises from a multiphase, non-equilibrium medium rather than a single homogeneous component. Comparison with photoionization and shock models shows that no single-component model within the explored parameter space can simultaneously reproduce both the strong neutral and high-ionization diagnostics, indicating that multiple physical conditions must coexist. We favor an interpretation in which shocks propagating through density inhomogeneities in the ejecta shape the observed morphology and excitation, while also considering alternative mechanisms linked to the central source, binary evolution, or interaction with an embedded object within the remnant. 
\end{abstract}

%% Keywords should appear after the \end{abstract} command. 
%% The AAS Journals now uses Unified Astronomy Thesaurus (UAT) concepts:
%% https://astrothesaurus.org
%% You will be asked to selected these concepts during the submission process
%% but this old "keyword" functionality is maintained in case authors want
%% to include these concepts in their preprints.
%%
%% You can use the \uat command to link your UAT concepts back its source.

%% From the front matter, we move on to the body of the paper.
%% Sections are demarcated by \section and \subsection, respectively.
%% Observe the use of the LaTeX \label
%% command after the \subsection to give a symbolic KEY to the
%% subsection for cross-referencing in a \ref command.
%% You can use LaTeX's \ref and \label commands to keep track of
%% cross-references to sections, equations, tables, and figures.
%% That way, if you change the order of any elements, LaTeX will
%% automatically renumber them.

%%%%%%%%%%%%%%%%%%%%%%%%%%%%%%%%%%%%%%%%%%%%%%%%%%%%%%%%%%%%%%%%%%%%%%%%%%%%%%%%%%
\section{Introduction} 

Oxygen-rich supernova remnants (O-rich SNRs) represent a rare subclass of young core-collapse remnants whose optical emission is dominated by metal-rich ejecta originating from the stellar interior, especially oxygen and other $\alpha$-group elements. First identified in objects such as Cassiopeia~A \citep[e.g.,][]{Chevalier1979}, these remnants are characterized by fast-moving ($\gtrsim$1000~km\,s$^{-1}$) knots of ejecta enriched in $\alpha$-elements (e.g., O, Ne, Mg, S) and showing little to no H and He. As a result, O-rich SNRs provide a direct observational probe of massive star nucleosynthesis and internal structure at the moment of explosion.

The optical emission from O-rich SNRs is generally interpreted as arising from radiative shocks driven by the reverse shock into dense ejecta clumps. The high metal abundances promote efficient cooling, leading to strong forbidden and permitted line emission in the post-shock cooling and recombination zones. In addition, shock-generated ultraviolet radiation can preionize or photoionize adjacent material, creating complex ionization structures that complicate spectral interpretation \citep[e.g.,][]{Itoh1981a,Itoh1986,Dopita1987,Sutherland1995}. The relative importance of shock excitation, precursor ionization, and photoionization remains an important issue in the interpretation of optical spectra from O-rich SNRs.

The SNR 1E~0102.2--7219 (hereafter E0102) is one of the brightest X-ray sources in the Small Magellanic Cloud (SMC). It was discovered by the \textit{Einstein Observatory} \citep{Seward1981} and subsequently classified as an O-rich SNR based on its optical spectrum \citep{Dopita1981}. The progenitor of E0102 is inferred to have been a massive star, with an estimated initial mass of roughly 15--40~M$_\odot$ \citep[e.g.,][]{Blair2000,Long2019}, that exploded a few thousands years ago \citep[e.g.,][]{Finkelstein2006}. Recent proper-motion measurements of oxygen-rich ejecta knots in the optical have refined the explosion age to 1738$\pm$175~yr and provided an improved estimate of the center of expansion \citep[CoE;][ see Figure~\ref{fig:wfm_nfm}~(a)]{Banovetz2021}.

High-resolution optical studies revealed an asymmetric distribution of fast-moving ejecta, enriched in products of carbon burning (O, Ne, Mg) and oxygen burning (S, Ar, Cl), arranged in an incomplete shell \citep{Vogt2010,Seitenzahl2018}. While the remnant is predominantly H-poor, localized fast-moving Balmer emission has been detected, implying that the progenitor may have retained part of its H envelope post-supernova explosion, possibly due to an interaction with a close companion star \citep{Seitenzahl2018}. Searches for a surviving companion using \textit{HST} data identified only one plausible candidate near the CoE of \citet{Banovetz2021}, though it remains unconfirmed due to its high implied transverse velocity \citep{Li2021}.

The forward shock of E0102 has been traced through optical coronal iron emission ([\ion{Fe}{14}] and [\ion{Fe}{11}] lines), which map the interaction of the blast wave with the surrounding medium and possibly a pre-existing stellar-wind cavity \citep{Vogt2017a}. No significant CO emission has been detected in the vicinity of E0102, indicating a relatively low-density ambient environment \citep{Alsaberi2024}.

Observations with the Multi Unit Spectroscopic Explorer in Wide Field Mode (MUSE WFM; \citealt{Bacon2010}) discovered a striking optical torus of \ion{O}{1} and \ion{Ne}{1} emission, and identified an X-ray point source near its geometric center \citep[][; see Figure~\ref{fig:wfm_nfm}~(a)]{Vogt2018}. Based on its X-ray properties and the absence of an optical counterpart, the point source was proposed to be a Central Compact Object (CCO) candidate, potentially associated with the surrounding optical torus \citep{Vogt2018}. Later X-ray studies offered contrasting interpretations: \citet{Hebbar2020} argued that the source is consistent with thermal emission from a neutron star, supporting its CCO nature, while \citet{Long2020} attributed the emission to localized hot ejecta.

The nature of this optical structure remains unclear: it has been interpreted as a torus associated with a putative CCO, but the spatial resolution of MUSE WFM was insufficient to resolve its internal structure. In particular, the physical mechanism and geometry responsible for the emission remain uncertain. In this work, we present new MUSE Narrow Field Mode with Adaptive Optics (NFM-AO) observations that provide an order-of-magnitude improvement in spatial resolution, allowing us to probe the morphology, kinematics, and excitation of the structure at sub-arcsecond scales.

%%%%%%%%%%%%%%%%%%%%%%%%%%%%%%%%%%%%%%%%%%%%%%%%%%%%%%%%%%%%%%%%%%%%%%%%%%%%%%%%%%
\section{Observations} \label{sec:obs}

%%%%%%%%%%%%%%%%%%%%%%%%%%%%%%%%%%%%%%%%%
\subsection{MUSE NFM-AO Observations}

The optical torus in E0102 was observed using MUSE in NFM with ground-layer adaptive optics (Program ID 0104.D--0092(A); P.I.: F.~P.~A.~Vogt). The NFM-AO configuration delivers a spatial sampling of 0.025\arcsec\,spaxel$^{-1}$ (spatial pixel) across a field of view of $\sim$7.5\arcsec$\times$7.5\arcsec, with wavelength coverage from 4750~\AA\ to 9350~\AA.

The observations were obtained over multiple service-mode visits, with individual pointings tiled into a mosaic pattern to sample the optical torus. Due to the constraints imposed by the AO system, particularly the availability and geometry of suitable reference sources for wavefront correction, the mosaic could not cover the full extent of the torus. The observing strategy was therefore optimized to maximize coverage of the brightest and most structurally informative regions while maintaining stable AO performance across this complex, spatially extended emission. A detailed observation log is provided in Appendix~\ref{app:nfm_obs}, Table~\ref{tab:obslog}.

%%%%%%%%%%%%%%%%%%%%%%%%%%%%%%%%%%%%%%%%%
\subsection{Data Reduction and Post-Processing}

Raw data were reduced using the automated European Southern Observatory (ESO) MUSE pipeline \citep{Weilbacher2020} within the \textsc{EsoReflex} environment \citep[ver.~2.11.5;][]{esoreflex}. Standard steps included bias subtraction, flat-fielding, wavelength and flux calibration, and illumination correction using ESO-provided calibration files. No dedicated sky frames were obtained due to the crowded, extended emission across the field, so no global sky subtraction was applied.

\begin{figure*}
\plotone{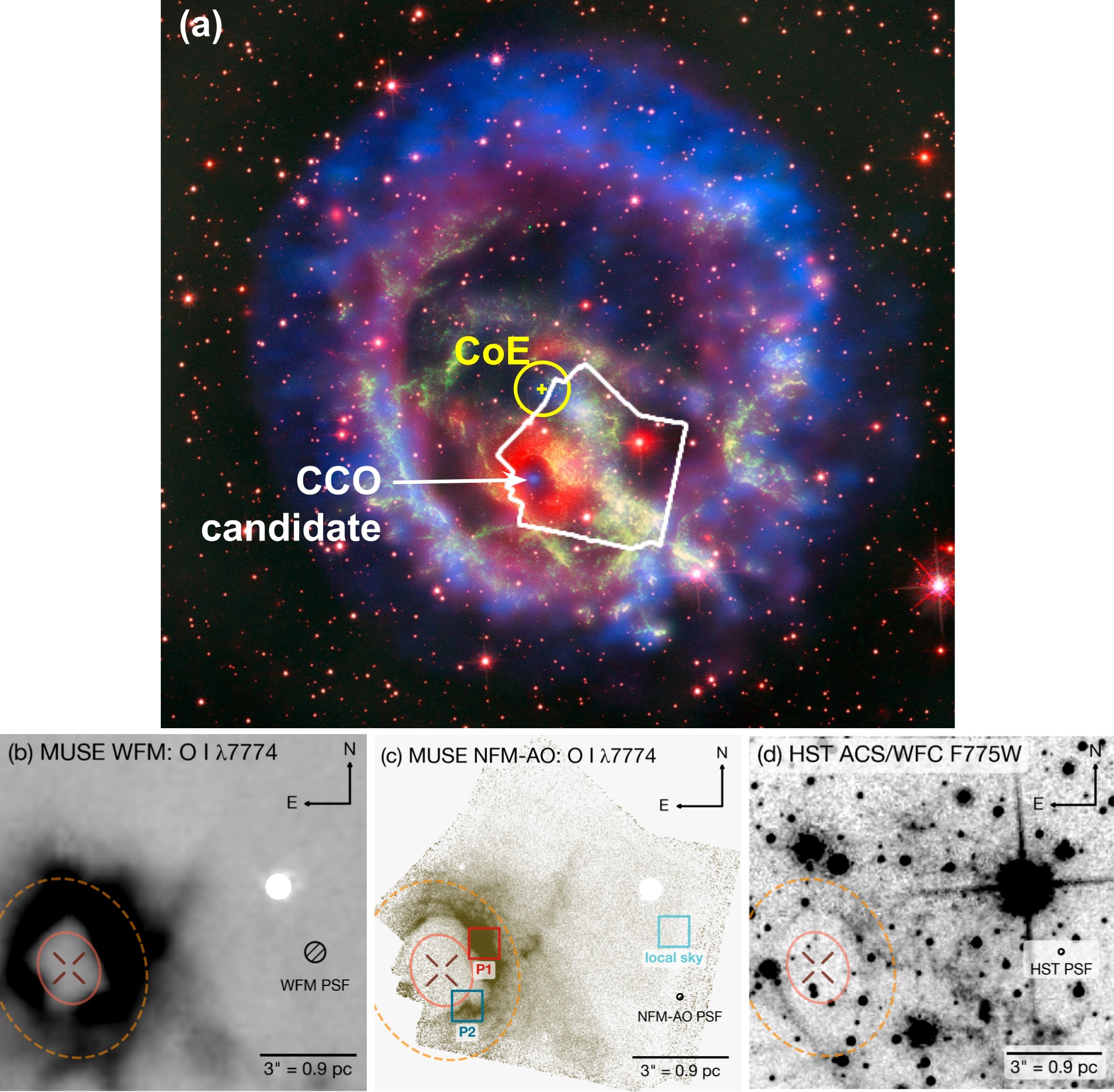}
\caption{
(a) Multiwavelength composite image of E0102 combining X-ray emission from \textit{Chandra} (blue and purple) with optical emission from VLT/MUSE (bright red) and \textit{HST} (dark red and green). The footprint of the MUSE NFM mosaic is shown in white and the CCO candidate identified by \citet{Vogt2018} is indicated by the white arrow. The center of expansion (CoE) determined from the proper-motion expansion of ejecta by \citet{Banovetz2021} is marked by the yellow cross, with the corresponding 1$\sigma$ uncertainty shown as the surrounding yellow circle. Image credits: X-ray (NASA/CXC/ESO/F.~Vogt et al.); Optical (ESO/VLT/MUSE \& NASA/STScI).
(b) Continuum-subtracted MUSE WFM \ion{O}{1} $\lambda$7774 cutout matched to the MUSE NFM field of view, integrated over $-50 \le v_{\text{los}} \le +450~\mathrm{km\,s^{-1}}$, highlighting the low-ionization torus. The solid and dashed ellipses trace the spatial extent of the structure used in \citet{Vogt2018}, and the dark-red crosshair marks the position of the X-ray source. The white circle indicates a masked bright foreground star. 
(c) Continuum-subtracted MUSE NFM-AO map of \ion{O}{1} $\lambda$7774 emission, integrated from $-50 \le v_{\text{los}} \le +450~\mathrm{km\,s^{-1}}$. The $1\arcsec \times 1\arcsec$ extraction regions (P1, P2) and a representative local sky region are shown. The purple line indicates the slice used to construct the position-velocity (PV) diagrams (see Section~\ref{subsec:cavity}). 
(d) \textit{HST} ACS/WFC F775W cutout matched to the MUSE NFM field of view, showing a sharp nebulosity at the same location as the structure outlined by the ellipse. 
}
\label{fig:wfm_nfm}
\end{figure*}

Individual reduced exposures were astrometrically aligned and combined into a final mosaic datacube. The final mosaic has dimensions of $467~\mathrm{spaxel}\times481$~spaxel, corresponding to an apparent area of 11.7\arcsec\,$\times$ 12.0\arcsec\ on the sky or a physical size of $3.5~\mathrm{pc}\times3.6$~pc at the SMC distance of 62~kpc \citep{Graczyk2020}. Post-processing with the \textsc{brutifus} package \citep{brutifus} refined the astrometry, applied foreground extinction corrections ($A_V=0.101$, $A_B=0.134$; following the extinction law of \citealt{Fitzpatrick1999} with $R_V=2.7$), and subtracted local continuum. In the absence of a dedicated sky exposure, a representative ``empty'' field region (see Figure~\ref{fig:wfm_nfm}~(c)) provided the local sky estimate for subtraction.

%%%%%%%%%%%%%%%%%%%%%%%%%%%%%%%%%%%%%%%%%%%%%%%%%%%%%%%%%%%%%%%%%%%%%%%%%%%%%%%%%%
\section{Data Analysis and Results} 
\label{sec:results}

Compared to the MUSE WFM data of \citet{Vogt2018}, the NFM-AO sampling of 0.025\arcsec~spaxel$^{-1}$ resolves filamentary substructures previously blended together (Figure~\ref{fig:wfm_nfm}~(b) and (c)). We detect permitted and forbidden transitions of oxygen, including \ion{O}{1} $\lambda\lambda$7774, 8446, 9262, 9266, [\ion{O}{1}]$\lambda\lambda$6300, 6363, [\ion{O}{2}]$\lambda\lambda$7320, 7330, and [\ion{O}{3}]$\lambda\lambda$4959, 5007, as well as faint emission from \ion{Ne}{1} $\lambda\lambda$6402, 6506, matching the lines identified in prior WFM observations (their Table~5).

%%%%%%%%%%%%%%%%%%%%%%%%%%%%%%%%%%%%%%%%%
\subsection{Morphology} 
\label{subsec:morphology}

\begin{figure*}
\plotone{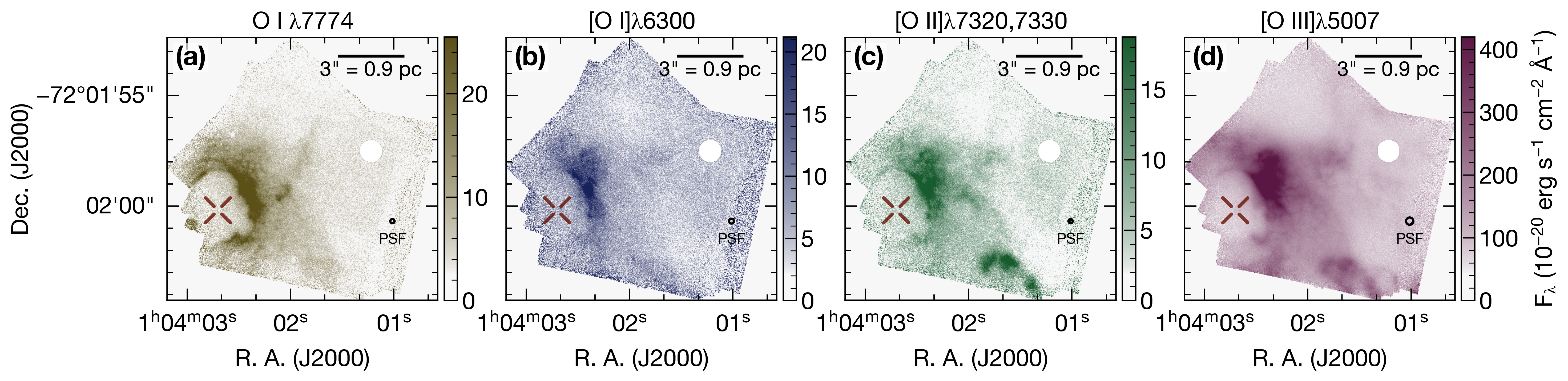}
\caption{
Integrated intensity maps of (a) \ion{O}{1} $\lambda$7774, (b) [\ion{O}{1}]$\lambda$6300, (c) [\ion{O}{2}]$\lambda$7320,7330, and (d) [\ion{O}{3}]$\lambda$5007, summed over $-50 \le v_{\text{los}} \le +450~\mathrm{km\,s^{-1}}$. The dark-red crosshair in each panel marks the position of the X-ray source, and the white circle indicates a masked bright foreground star. The MUSE NFM-AO PSF is shown in each panel, with measured values of $0.096\arcsec$, $0.110\arcsec$, $0.096\arcsec$, and $0.164\arcsec$ for \ion{O}{1}, [\ion{O}{1}], [\ion{O}{2}], and [\ion{O}{3}], respectively.
}
\label{fig:int_map}
\end{figure*}

Figure~\ref{fig:int_map} presents integrated intensity maps of the strongest lines from each oxygen ionization state: \ion{O}{1} $\lambda$7774, [\ion{O}{1}]$\lambda$6300, [\ion{O}{2}]$\lambda$7320,7330, and [\ion{O}{3}]$\lambda$5007, integrated over $-50 \le v_{\text{los}} \le +450~\mathrm{km\,s^{-1}}$, capturing the full torus emission. In contrast, [\ion{O}{3}] emission across the entire remnant spans $-3500$~km\,s$^{-1}$ to $+5000$~km\,s$^{-1}$ \citep{Vogt2017b}. The dark-red crosshair in each panel marks the CCO candidate (R.A: 01$^\mathrm{h}$04$^\mathrm{m}$02.7$^\mathrm{s}$, Dec: -72\arcdeg02\arcmin00.2\arcsec [J2000]; \citealt{Vogt2018}).

The permitted \ion{O}{1} $\lambda$7774 emission traces a sharply defined inner edge with filamentary substructures and a more diffuse outer extent. The sharp edge is also evident in \textit{HST} broadband imaging through multiple filters, for example in F775W ($\lambda_\mathrm{cen}=7763$~\AA, W$_\mathrm{eff}=1380$~\AA; Figure~\ref{fig:wfm_nfm}~(d); see also Figure~5 of \citealt{Vogt2018}). Forbidden-line emission follows a similar large-scale morphology as the \ion{O}{1} emission, including the pronounced inner edge, but extend further westward. Note that the AO correction is more effective at longer wavelengths, leading to a sharper point spread function (PSF) in the red part of the datacube. As a result, shorter-wavelength emission such as [\ion{O}{3}]$\lambda$5007 appears relatively more diffuse.

\begin{figure*}
\plotone{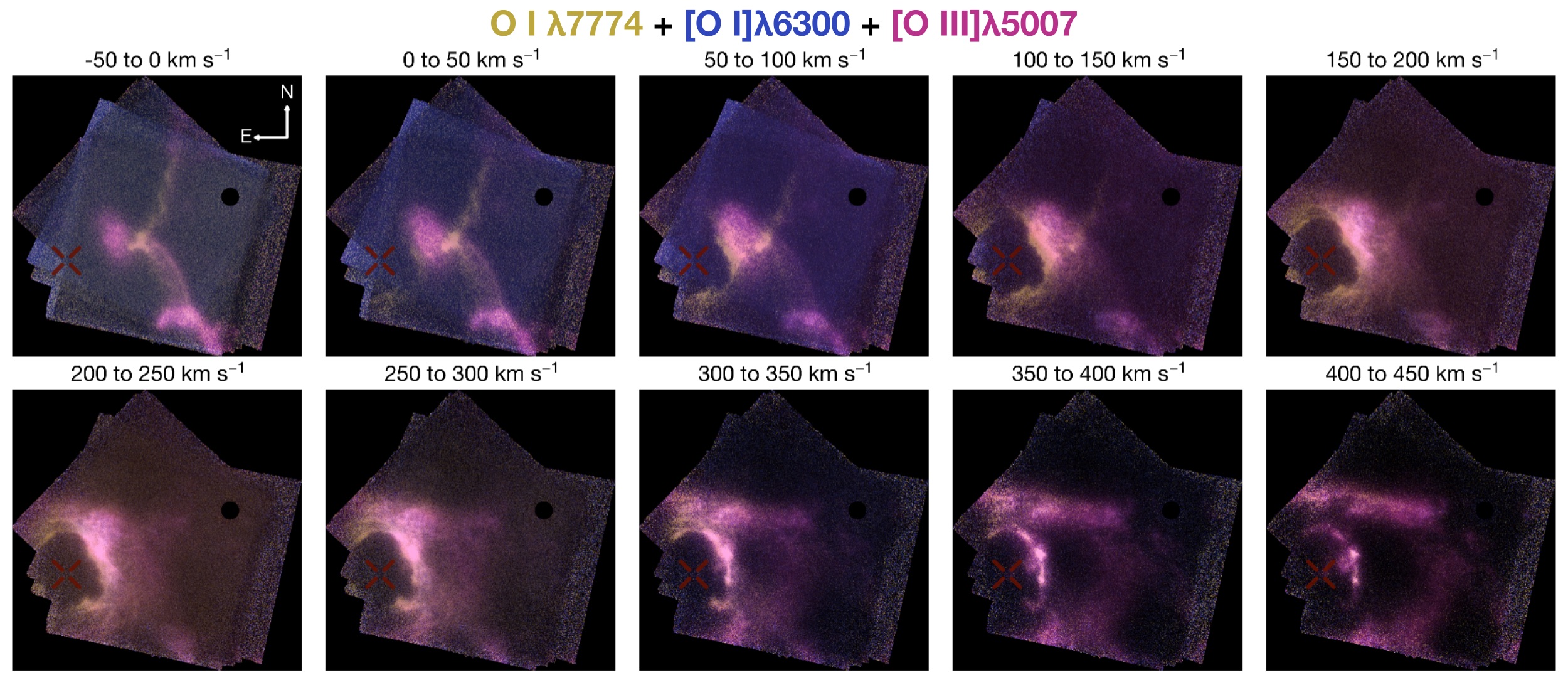}
\caption{Channel maps of a three-color composite: \ion{O}{1} $\lambda7774$ (yellow), [\ion{O}{1}]$\lambda6300$ (blue), and [\ion{O}{3}]$\lambda5007$ (magenta), constructed in 50~km\,s$^{-1}$ bins from $-50$ to $450~\mathrm{km\,s^{-1}}$. The orientation on the sky is indicated in the upper-left panel. The dark-red crosshair marks the position of the X-ray source, and the black circle indicates a masked foreground star. 
}
\label{fig:chanmaps_composites}
\end{figure*}

Channel maps in Figure~\ref{fig:chanmaps_composites} combine \ion{O}{1} $\lambda$7774 (yellow), [\ion{O}{1}]$\lambda$6300 (blue), and [\ion{O}{3}]$\lambda$5007 (magenta) in 50~km\,s$^{-1}$ bins from $-50~\mathrm{km\,s^{-1}}$ to $450~\mathrm{km\,s^{-1}}$. Single-line versions are shown in Appendix~\ref{app:chanmap} (Figure~\ref{fig:chanmap_all}). Across this velocity range, emission varies smoothly without discrete kinematic breaks or clear red/blue segregation expected from a toroidal geometry. The sharp inner edge, most prominently traced by \ion{O}{1} emission, emerges first on the south at $+50$ to $+100$~km\,s$^{-1}$, brightens and encircles the structure, then fades at $+350$ to $+400$~km~s$^{-1}$. At $v_{\mathrm{los}} \gtrsim 300$~km~s$^{-1}$, [\ion{O}{3}] fills the interior (material receding from us in projection), while [\ion{O}{1}] emission broadly agrees with \ion{O}{1} emission. At velocities between $-50$ and $+100$~km\,s$^{-1}$, however, the [\ion{O}{1}] maps suffer strong residual sky contamination.

\begin{figure*}
\plotone{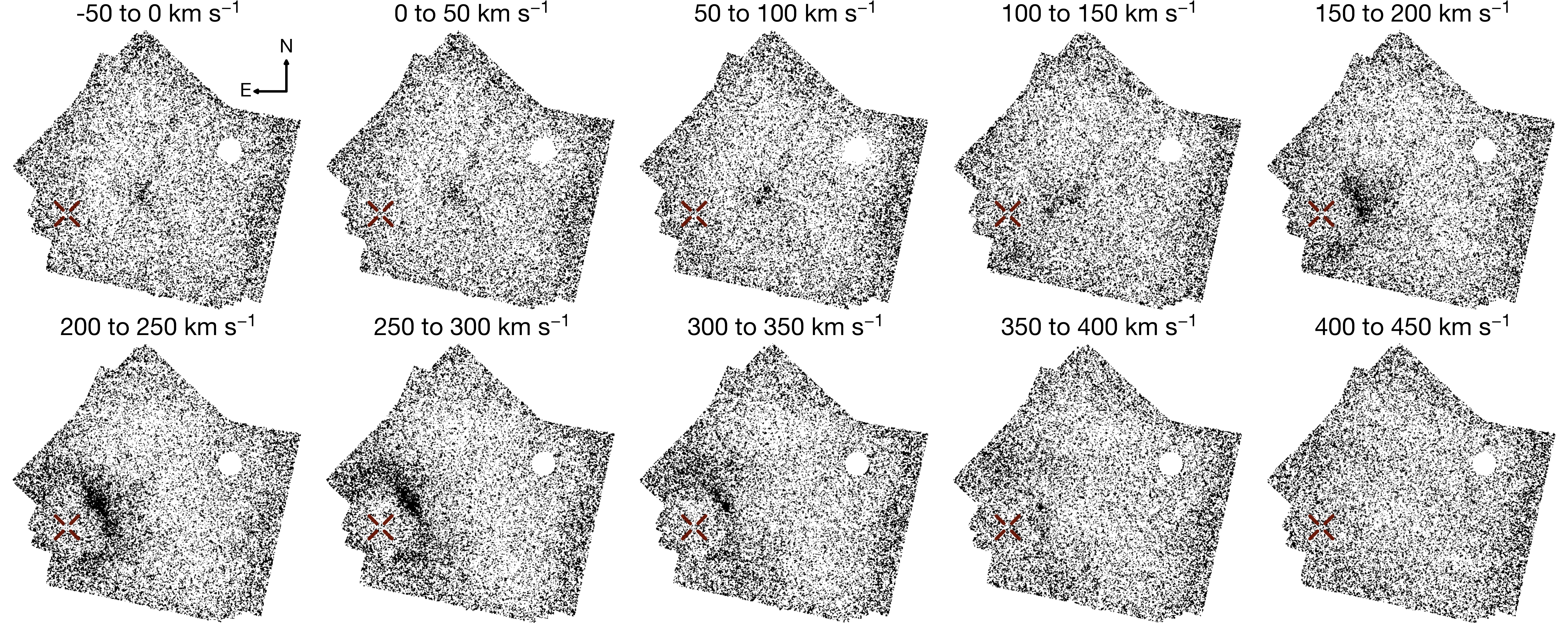}
\caption{
Channel maps of \ion{Ne}{1} $\lambda$6402 emission in 50~km\,s$^{-1}$ bins from $-50$ to $+450~\mathrm{km\,s^{-1}}$. A Gaussian smoothing has been applied to enhance low surface brightness structures. The orientation on the sky is indicated in the upper-left panel. The dark-red crosshair marks the position of the X-ray source, and the white circle indicates a masked foreground star.  
}
\label{fig:chanmap_ne1}
\end{figure*}

Although significantly fainter than the \ion{O}{1} $\lambda$7774 emission, \ion{Ne}{1} $\lambda$6402 is clearly detected spatially (Figure~\ref{fig:chanmap_ne1}). Across the velocity channels, the \ion{Ne}{1} emission broadly follows the same spatial distribution as \ion{O}{1}, with the clearest signal observed between $+150$ and $+300~\mathrm{km\,s^{-1}}$, confirming a co-spatial origin despite lower signal-to-noise ratio (S/N).

%%%%%%%%%%%%%%%%%%%%%%%%%%%%%%%%%%%%%%%%%
\subsection{Spectral Analysis} 
\label{subsec:spectral}

We extracted the integrated spectra from two representative 1\arcsec$\times$1\arcsec\ regions (P1 and P2; Figure~\ref{fig:wfm_nfm}~(c)). Local sky subtraction used a nearby emission-free region, verified against standard sky lines \citep{Osterbrock1996}. Emission lines were fitted with Gaussian profiles. For transitions within the same ion species, we tied the velocity centroids and intrinsic widths, applying fixed theoretical flux ratios where appropriate (i.e., [\ion{O}{3}] $F_{5007}/F_{4959}=2.98$). To derive the intrinsic full width at half maximum (FWHM$_\mathrm{int}$), instrumental broadening was corrected using:
\begin{equation}
    \mathrm{FWHM}_{\mathrm{int}} = \sqrt{\mathrm{FWHM}_{\mathrm{obs}}^2 - \mathrm{FWHM}_{\mathrm{inst}}^2}.
\end{equation}
For permitted \ion{O}{1} lines, all shared a common centroid, but we allowed independent widths due to the systematically broader $\lambda$7774 profile. The resulting line parameters are summarized in Table~\ref{tab:line_measurements}.

\begin{table*}
\centering
\caption{Emission-line measurements from MUSE NFM-AO spectra extracted from two $1\arcsec \times 1\arcsec$ regions (P1 and P2; Figure~\ref{fig:wfm_nfm}~(c)). For each transition, we report the surface brightness, line-of-sight velocity, and intrinsic FWHM. 
}
\vspace{-5pt}
\begin{tabular}{llcccccc}
\toprule
\addlinespace[1.5pt]
\toprule
  &  & \multicolumn{3}{c}{P1} & \multicolumn{3}{c}{P2} \\
\cmidrule(lr){3-5}\cmidrule(lr){6-8}
~Line & $\lambda_{\text{rest}}$ & S ($\times10^{-16}$)
& $v_{\text{los}}$ & FWHM$_{\text{int}}$ & S ($\times10^{-16}$) & $v_{\text{los}}$ & FWHM$_{\text{int}}$ \\
  & (\AA) & 
  {\footnotesize erg\,s$^{-1}$\,cm$^{-2}$\,arcsec$^{-2}$} & {\footnotesize km\,s$^{-1}$} &
  {\footnotesize km\,s$^{-1}$} & {\footnotesize erg\,s$^{-1}$\,cm$^{-2}$\,arcsec$^{-2}$} & 
  {\footnotesize km\,s$^{-1}$} & {\footnotesize km\,s$^{-1}$} \\
\midrule
~[\ion{O}{3}]$\lambda$4959 & 4958.91 & $17.15 \pm 0.03$ & $230.9 \pm 0.3$ & $210.0 \pm 0.6$ 
& $2.63 \pm 0.02$ & $276.7 \pm 0.6$ & $43.5 \pm 5.0$ \\ 
~[\ion{O}{3}]$\lambda$5007 & 5006.84 & $51.10 \pm 0.10$ & $230.9 \pm 0.3$ & $210.0 \pm 0.6$ 
& $7.84 \pm 0.06$ & $276.7 \pm 0.6$ & $43.5 \pm 5.0$ \\ 
~[\ion{O}{1}]$\lambda$6300 & 6300.30 & $2.73 \pm 0.01$ & $260.4 \pm 0.4$ & $197.0 \pm 1.1$ 
& $0.559 \pm 0.012$ & $269.5 \pm 1.4$ & $146.6 \pm 5.1$ \\
~[\ion{O}{1}]$\lambda$6363 & 6363.78 & $0.894 \pm 0.006$ & $260.4 \pm 0.4$ & $197.0 \pm 1.1$ 
& $0.188 \pm 0.007$ & $269.5 \pm 1.4$ & $146.6 \pm 5.1$ \\ 
~\ion{Ne}{1} $\lambda$6402 & 6402.25 & $0.404 \pm 0.006$ & $258.0 \pm 1.2$ & $150.1 \pm 3.8$ 
& -- & -- & -- \\
~\ion{Ne}{1} $\lambda$6506 & 6506.53 & $0.220 \pm 0.005$ & $258.0 \pm 1.2$ & $150.1 \pm 3.8$ 
& -- & -- & -- \\ 
~[\ion{O}{2}]$\lambda$7320 & 7319.92\tablenotemark{a} & $3.14 \pm 0.01$ & $238.5 \pm 0.1$ & $198.4 \pm 0.4$ & $0.800 \pm 0.007$ & $264.4 \pm 0.5$ & $169.6 \pm 1.6$ \\
~[\ion{O}{2}]$\lambda$7330 & 7330.19\tablenotemark{a} & $2.36 \pm 0.01$ & $238.5 \pm 0.1$ & $198.4 \pm 0.4$ & $0.567 \pm 0.006$ & $264.4 \pm 0.5$ & $169.6 \pm 1.6$ \\
~\ion{O}{1} $\lambda$7774 & 7774.17\tablenotemark{a} & $5.59 \pm 0.00$ & $215.7 \pm 0.1$ & $190.1 \pm 0.2$ & $2.38 \pm 0.01$ & $206.0 \pm 0.2$ & $171.1 \pm 0.6$ \\
~\ion{O}{1} $\lambda$8446 & 8446.36\tablenotemark{a} & $2.01 \pm 0.01$ & $215.7 \pm 0.1$ & $108.1 \pm 0.5$ & $1.09 \pm 0.01$ & $206.0 \pm 0.2$ & $68.9 \pm 1.2$ \\ 
~\ion{O}{1} $\lambda$9262 & 9262.67\tablenotemark{a} & $1.05 \pm 0.01$ & $215.7 \pm 0.1$ & $144.3 \pm 2.2$ & $0.383 \pm 0.013$ & $206.0 \pm 0.2$ & $89.8 \pm 6.8$ \\
~\ion{O}{1} $\lambda$9266 & 9266.01\tablenotemark{a} & $1.50 \pm 0.01$ & $215.7 \pm 0.1$ & $144.3 \pm 2.2$ & $0.490 \pm 0.014$ & $206.0 \pm 0.2$ & $89.8 \pm 6.8$ \\
\bottomrule
\end{tabular}
\tablenotetext{a}{\parbox[t]{0.9\textwidth}{\raggedright multiplets}}
\label{tab:line_measurements}
\end{table*}

The measured line fluxes and kinematics agree well with the MUSE WFM values reported by \citep{Vogt2018}, confirming that the NFM-AO observations resolve the same physical structure. The lower S/N per spaxel in the NFM-AO data, particularly for \ion{Ne}{1} emission, simply reflects the dramatically finer spatial sampling rather than any intrinsic difference in emission strength.

Line-of-sight velocities show good coherence across species, spanning $\sim$216 to $\sim$260~km~s$^{-1}$ in P1 and $\sim$206 to $\sim$277~km~s$^{-1}$ in P2. Most forbidden lines and the permitted \ion{O}{1} $\lambda$7774 exhibit comparable intrinsic widths of $\sim$190--210~km~s$^{-1}$, suggesting a shared kinematic origin (the sole exception being the unusually narrow [\ion{O}{3}] in P2). However, the permitted \ion{O}{1} lines reveal systematic differences: $\lambda$7774 is significantly broader than $\lambda$8446 in both regions, indicating that the former samples more extended or turbulent neutral material.

Purely thermal Doppler broadening at these widths would require temperatures of $\sim$10$^7$~K, which are not possible with neutral species like \ion{O}{1} and \ion{Ne}{1}. Non-thermal motions must therefore dominate, whether from turbulence, unresolved velocity gradients, or line-of-sight superposition of multiple components. The smooth line profiles, lacking secondary peaks or clear substructure, point toward bulk gas motions and projection geometry as the primary broadening mechanisms within the extraction aperture.

%%%%%%%%%%%%%%%%%%%%%%%%%%%%%%%%%%%%%%%%%
\subsection{\ion{O}{1} Kinematics} \label{subsec:kinematics}

\begin{figure*}
\plotone{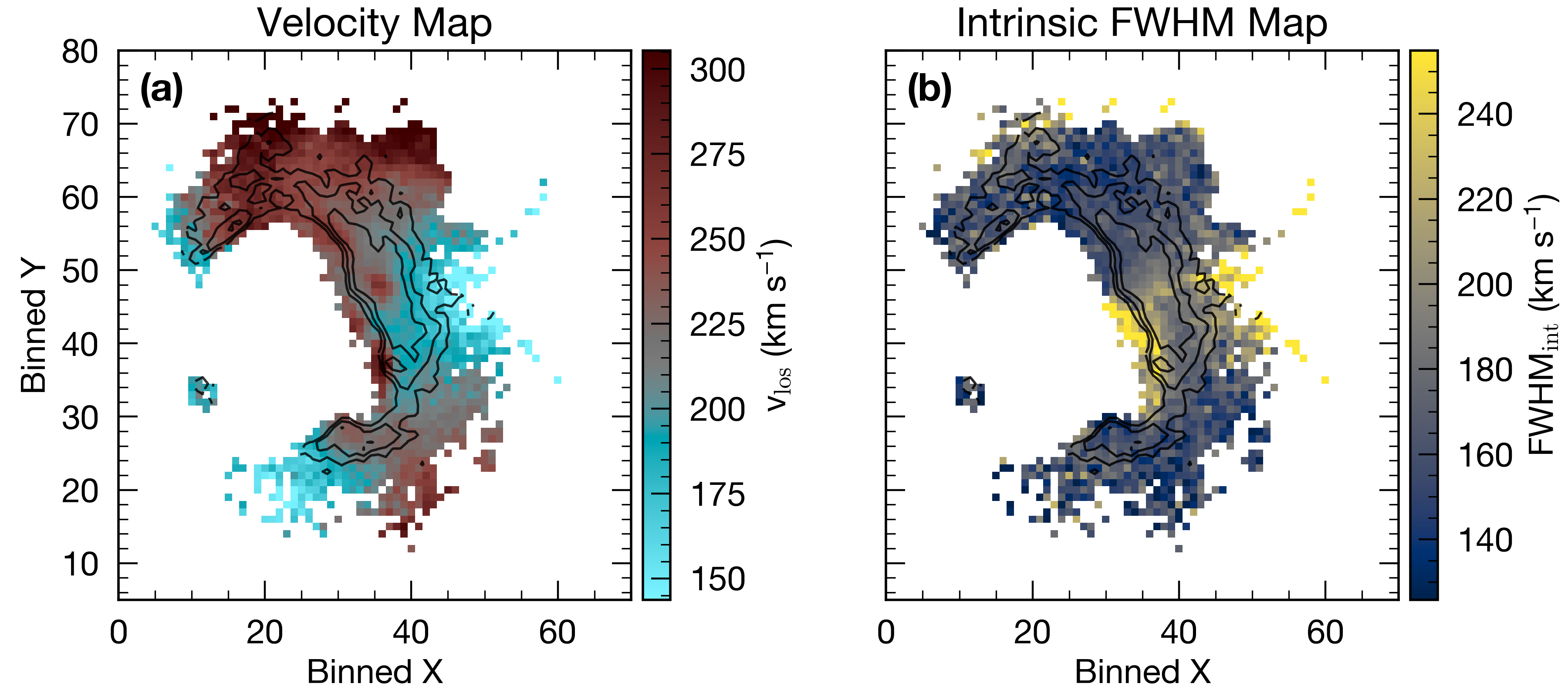}
\caption{(a) Line-of-sight velocity map of \ion{O}{1} $\lambda\lambda$7774,8446, derived from spaxel-by-spaxel Gaussian fitting, with \ion{O}{1} $\lambda$7774 flux contours overlaid to trace the emission morphology. (b) Intrinsic FWHM map of \ion{O}{1} $\lambda7774$, corrected for instrumental broadening, with \ion{O}{1} $\lambda$7774 flux contours overlaid.
Both maps are constructed from data spatially binned by a factor of 4 and include only spaxels with $\mathrm{S/N}\geq5$. The contours correspond to the top 25\% of the \ion{O}{1} $\lambda7774$ flux distribution, highlighting the brightest regions of the emission.
}
\label{fig:vel_fwhm_map}
\end{figure*}

We performed spaxel-by-spaxel Gaussian fits to the \ion{O}{1} $\lambda\lambda$7774,8446 lines after 4$\times$ spatial binning, retaining only spaxels with S/N $\ge 5$. The resulting velocity and intrinsic FWHM maps are shown in Figure~\ref{fig:vel_fwhm_map}, with the divergent points of the colormaps set to match the P1 $\lambda$7774 values.

The velocity map confirms a large-scale velocity gradient reported in \citet{Vogt2018}. While this gradient was interpreted as arising from the projected geometry of a tilted torus, the higher-spatial-resolution NFM-AO data, combined with the channel maps (Figure~\ref{fig:chanmaps_composites}; see also Figure~\ref{fig:chanmap_all}), reveal a much smoother, continuous velocity evolution rather than a discrete rotation or expansion signature expected from a toroidal geometry. Line widths remain systematically broad ($\sim$140--240~km\,s$^{-1}$) throughout the structure, with particularly enhanced values along the western inner edge near P1.

The large-scale velocity gradient and spatially varying intrinsic line widths suggest that the observed kinematics likely reflect a combination of projection effects and unresolved internal motions within the structure, including turbulence or superposition of multiple clumps along the line of sight.

%%%%%%%%%%%%%%%%%%%%%%%%%%%%%%%%%%%%%%%%%
\subsection{The Cavity}
\label{subsec:cavity}

The morphology of the \ion{O}{1} emission, as shown in Figures~\ref{fig:int_map} and \ref{fig:chanmaps_composites}, reveals an emission structure characterized by a sharply defined inner edge and filamentary emission extending outward. While this structure has previously been described as a torus \citep{Vogt2018}, the presence of a sharp edge is not naturally expected in a smooth, toroidal distribution. Instead, it points to a cavity geometry, in which emission is concentrated at a sharp edge surrounding a region of reduced intensity toward smaller radii.

\begin{figure*}
\plotone{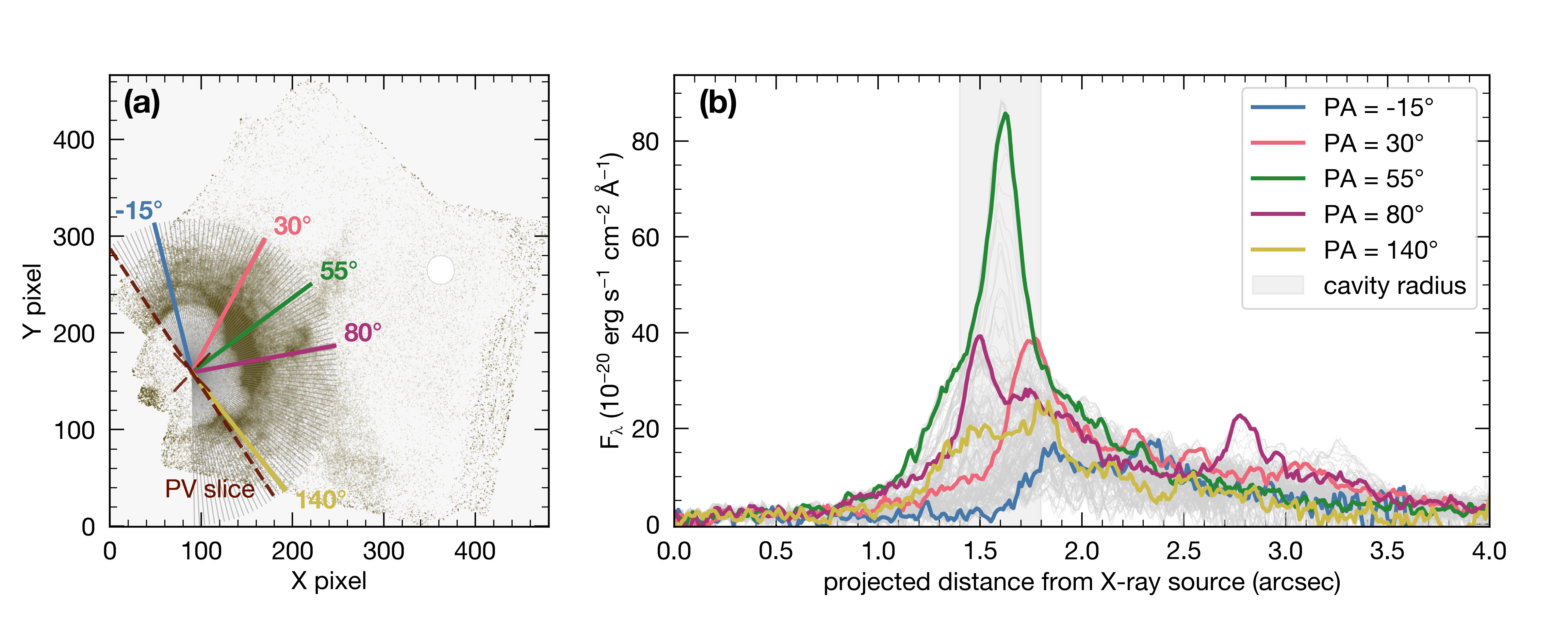}
\caption{(a) Integrated \ion{O}{1} $\lambda7774$ intensity map with radial cuts spanning a range of position angles. All sampled position angles are shown in grey, with selected directions highlighted to illustrate representative profiles. The dashed line indicates the orientation of the position--velocity (PV) slice. The X-ray source position is marked by the red crosshair.
(b) Radial intensity profiles as a function of projected distance from the X-ray source. Individual profiles are shown in grey, while selected position angles are highlighted in color. The shaded region marks the characteristic radius of the cavity $\sim$1.4\arcsec--1.8\arcsec.
}
\label{fig:radial_profile}
\end{figure*}

To examine the radial structure of the emission, we extract intensity profiles as a function of projected distance from the X-ray source over a range of position angles (PA; see Figure~\ref{fig:radial_profile}). The profiles show near zero emission at small radii ($\lesssim$0.5\arcsec), followed by a sharp rise at $\sim$1.4\arcsec--1.8\arcsec, and a more extended decline toward larger radii. This behavior differs from that expected for a smooth torus, which would produce a more symmetric radial profile with a gradual rise and fall about its characteristic radius (i.e., a Gaussian profile). Instead, the observed profiles show a sharp edge at a characteristic radius, consistent with the boundary of a cavity rather than a smoothly distributed torus.

\begin{figure}[ht!]
\plotone{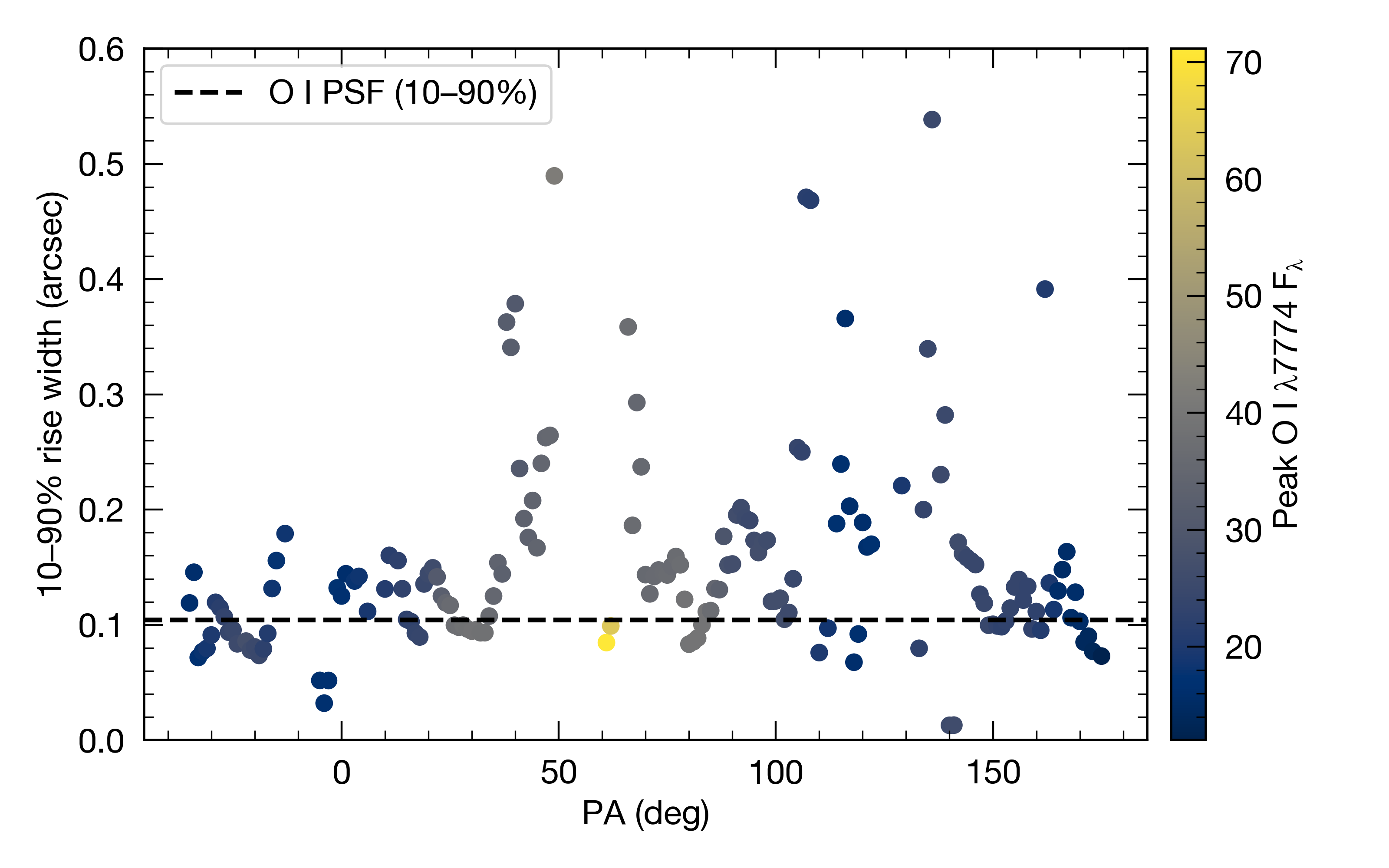}
\caption{
10\%--90\% intensity rise width of the \ion{O}{1} $\lambda7774$ radial profiles as a function of position angle. The dashed line indicates the corresponding width of the instrumental PSF. The color scale shows the peak intensity of each profile. The measured widths are generally close to the PSF, indicating a sharp, spatially confined edge.
}
\label{fig:edge_width_vs_pa}
\end{figure}

The sharpness of this transition is quantified in Figure~\ref{fig:edge_width_vs_pa}, which shows the 10\%--90\% rise width of the \ion{O}{1} radial profiles as a function of PA. For each profile, this width is defined as the radial distance over which the intensity increases from 10\% to 90\% of its peak value across the edge. In most directions, the measured widths are comparable to or only modestly larger than the instrumental point-spread function, indicating that the edge is resolved but remains intrinsically narrow.

\begin{figure*}
\plotone{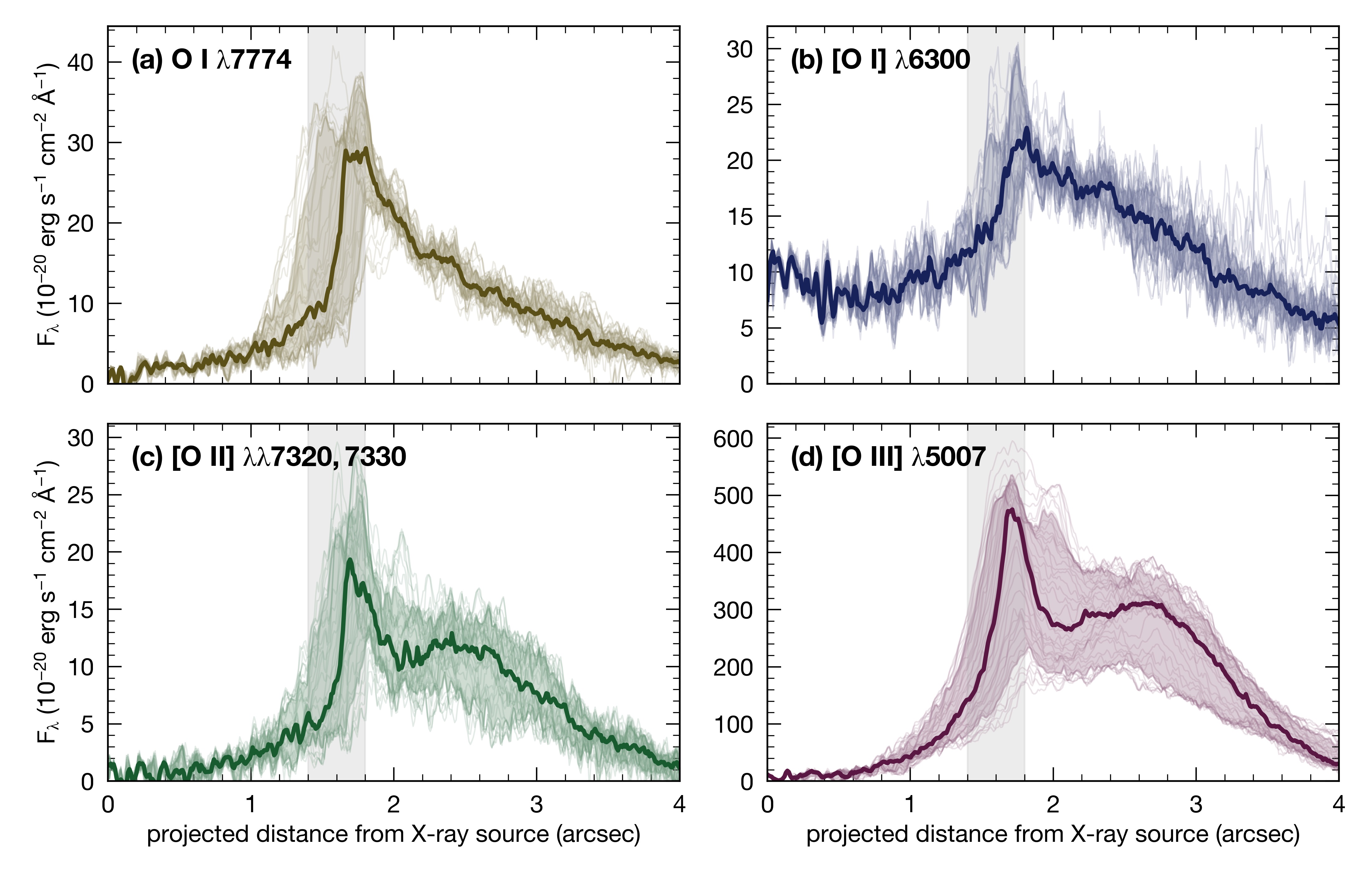}
\caption{
Radial intensity profiles for (a) \ion{O}{1}, (b) [\ion{O}{1}], (c) [\ion{O}{2}], and (d) [\ion{O}{3}] emission, showing the median profile (solid line) and the 16$^\text{th}$--84$^\text{th}$ percentile range (shaded region) over position angles between $10^\circ$ and $50^\circ$. The shaded vertical band marks the characteristic radius of the edge. 
}
\label{fig:radial_profiles_lines}
\end{figure*}

The radial profile behavior across different emission lines further clarifies the structure. Figure~\ref{fig:radial_profiles_lines} shows the profiles correspond to PA between $10^\circ$ and $50^\circ$, where the emission is strongest for \ion{O}{1}, [\ion{O}{1}], [\ion{O}{2}], and [\ion{O}{3}]. All lines exhibit a similar sharp rise, most pronounced in \ion{O}{1}, while higher-ionization emission such as [\ion{O}{3}] extends more gradually outward. This systematic variation indicates a multiphase structure, in which neutral and ionized gas coexist and trace different regions of the same geometry.

\begin{figure}
\plotone{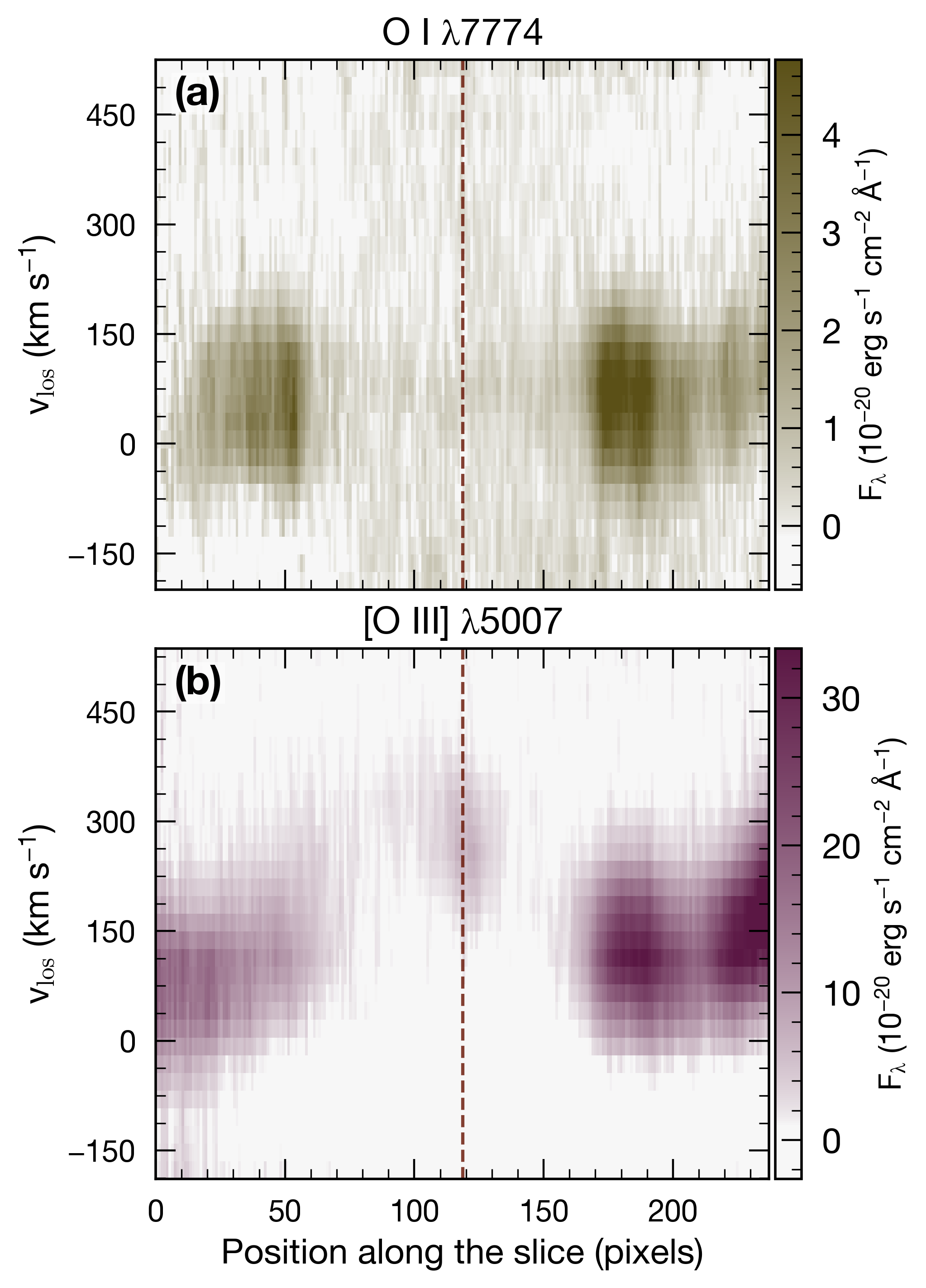}
\caption{PV diagrams of (a) \ion{O}{1} $\lambda7774$ and (b) [\ion{O}{3}]$\lambda5007$ (bottom), extracted along a slice that crosses the structure and passes through the X-ray source (shown in Figure~\ref{fig:radial_profile}). The red dashed line marks the position of the X-ray source along the slice. The \ion{O}{1} emission shows two bright concentrations at opposite sides of the slice, corresponding to the projected boundaries of the cavity-like structure. The [\ion{O}{3}] emission shows a similar pair of bright concentrations, but also exhibits a distinct higher-velocity component within the projected interior, particularly at $\sim$200--350~km\,s$^{-1}$.
}
\label{fig:pv_diagram}
\end{figure}

The sharp edge identified in the radial profiles is further clarified by the position--velocity (PV) structure. Figure~\ref{fig:pv_diagram} shows the PV diagrams of \ion{O}{1} $\lambda7774$ and [\ion{O}{3}]$\lambda5007$ extracted along a slice that crosses the structure and passes through the X-ray source (dashed line in Figure~\ref{fig:radial_profile}).

The \ion{O}{1} PV diagram (Figure~\ref{fig:pv_diagram}~(a)) shows two bright concentrations on opposite sides of the slice, separated by a region of low emission. These components trace the edges of the structure in projection. The separation between the peaks yields a characteristic radius of $\sim$1.6\arcsec\ (Appendix~\ref{app:radius}), consistent with the radius inferred from the radial profiles.

While the [\ion{O}{3}] PV diagram (Figure~\ref{fig:pv_diagram}~(b)) mirrors this morphology, it also reveals a distinct higher-velocity component filling the cavity interior at positions $\sim$100--130 pixels and velocities of $\sim$200--350~km\,s$^{-1}$, representing a receding gas seen in projection through the cavity.

The morphology, radial structure, edge sharpness, and kinematic behavior of the emission are not consistent with a geometrically thin torus. A toroidal ring would produce symmetric radial profiles and emission confined to a relatively narrow radial range. Instead, the observations are best explained by a cavity structure, in which emission arises from a sharply defined edge.

%%%%%%%%%%%%%%%%%%%%%%%%%%%%%%%%%%%%%%%%%
\subsection{Spatially Resolved Line Ratios} 
\label{subsec:lineratios}

\begin{figure}
\plotone{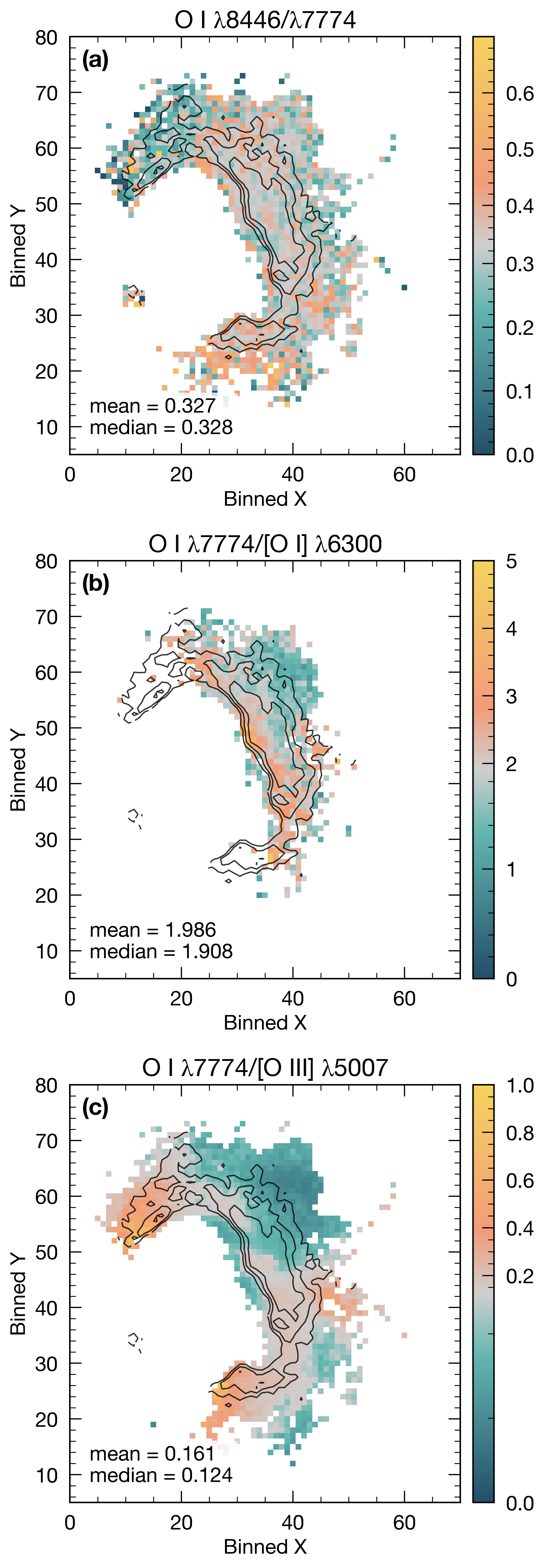}
\caption{Spatially resolved line-ratio maps of the optical torus in E0102. 
(a) \ion{O}{1} $\lambda8446/\lambda7774$, tracing the variations within permitted \ion{O}{1} emission. 
(b) \ion{O}{1} $\lambda7774$/[\ion{O}{1}]$\lambda6300$, tracing the recombination-favored and collisionally excited neutral oxygen. 
(c) \ion{O}{1} $\lambda7774$/[\ion{O}{3}]$\lambda5007$, tracing the transition between low- and high-ionization gas. 
All maps are derived from spaxel-by-spaxel Gaussian fits to data binned by a factor of 4, with only spaxels of $\mathrm{S/N} \geq 5$ in both lines included. Black contours trace the top 25\% of the \ion{O}{1} $\lambda7774$ flux distribution. }
\label{fig:linerat_map}
\end{figure}

To investigate the excitation conditions, we constructed spatially resolved line-ratio maps using the spaxel-by-spaxel Gaussian fits (Figure~\ref{fig:linerat_map}). The datacube was spatially binned by a factor of 4, and only spaxels with $\mathrm{S/N} \geq 5$ in both lines were retained. Regions where the ratio cannot be reliably measured are masked. The black contours correspond to the top 25\% of the \ion{O}{1} $\lambda7774$ flux distribution.

The \ion{O}{1} $\lambda8446/\lambda7774$ ratio is a well-known diagnostic of the excitation mechanism of neutral oxygen emission, distinguishing between recombination-dominated and Ly$\beta$ fluorescence-dominated regimes. In fluorescence-dominated conditions, values of $\lambda8446/\lambda7774 > 1$ are typically expected, whereas lower values ($\lesssim$0.5) are more consistent with recombination. The observed ratio is relatively uniform across the structure (see Figure~\ref{fig:linerat_map}~(a)), with a median value of $\sim$0.33, indicating that fluorescence is not the dominant excitation mechanism and that the excitation conditions governing the permitted \ion{O}{1} transitions do not vary strongly across the emitting region.

In contrast, the \ion{O}{1} $\lambda7774$/[\ion{O}{1}]$\lambda6300$ ratio provides a qualitative tracer of the relative importance of recombination and collisionally excited neutral oxygen emission. Figure~\ref{fig:linerat_map}~(b) shows that there are clear spatial variations in \ion{O}{1} $\lambda7774$/[\ion{O}{1}]$\lambda6300$ ratio, with enhanced values concentrated along the cavity edge, and lower values in the more diffuse surrounding emission. This pattern indicates that the excitation conditions are not homogeneous, with recombination contributing more strongly along the cavity edge, while collisional excitation dominates more in the outward zones. We note, however, that this ratio also depends on density, temperature, ionization structure, and optical depth, and the two lines may not arise from exactly the same physical zones; it should therefore be interpreted as a qualitative diagnostic.

The \ion{O}{1} $\lambda7774$/[\ion{O}{3}]$\lambda5007$ ratio (Figure~\ref{fig:linerat_map}~(c)) highlights the spatial transition between low- and high-ionization gas. Low values correspond to regions dominated by highly ionized material traced by [\ion{O}{3}], while higher values trace regions where neutral or partially ionized gas is more prominent. The map shows that \ion{O}{1}-dominated emission traces the sharp edge of the cavity, whereas [\ion{O}{3}] emission extends outward, consistent with the trends seen in the channel maps (Figure~\ref{fig:chanmaps_composites}).

These diagnostics indicate that there are transitions in physical conditions across the structure, where partially ionized, recombining gas is confined within a more highly ionized and diffuse medium. Shocks propagating through an inhomogeneous medium under approximately constant ram pressure provide a straightforward explanation for this. The denser regions along the cavity edge experience slower shocks, leading to lower post-shock temperatures that favor recombination, enhancing permitted \ion{O}{1} relative to collisionally-excited [\ion{O}{1}]. Conversely, lower-density regions in the more diffuse outer zones are processed by faster shocks, resulting in higher temperatures that boost collisionally-excited emission while suppressing recombination.

This scenario explains the observed morphology, in which the recombination-dominated \ion{O}{1} emission appears brighter and concentrated along the cavity edge, while the more diffuse [\ion{O}{1}] and [\ion{O}{3}] emission traces lower-density regions extending outward. The emission is therefore best understood as a structured, multiphase region shaped by shocks propagating through density inhomogeneities in the ejecta.

%%%%%%%%%%%%%%%%%%%%%%%%%%%%%%%%%%%%%%%%%
\subsection{Comparison with Photoionization and Shock Models Predictions} 
\label{subsec:photo_shock_model}

We compared the observed line ratios with \textsc{Mappings} \citep[v.5.2.0;][]{mappings} photoionization and shock models. For the photoionization models, we considered two ionizing sources: (i) a blackbody with temperature consistent with the inferred X-ray source in \citet{Vogt2018} and \citet{Hebbar2020}, and (ii) thermal emission from a hot O- and Ne-rich ejecta knot as suggested by \citet{Long2020}.

The shock models covered preshock densities $n_{\rm H} = [10-1000]~\mathrm{cm^{-3}}$, velocities of $20 < v_s < 200~\mathrm{km~s^{-1}}$, and self-consistently include both the post-shock cooling region and the photoionized precursor. The precursor arises from ionizing UV radiation generated in the post-shock gas, which propagates upstream and heats and ionizes the incoming material \citep{Sutherland1995,Sutherland2017}. As a result, the predicted emission is naturally separated into contributions from the post-shock cooling and recombination zone (``shock-only'') and the upstream photoionized gas (``precursor'').

We explored both SMC abundances \citep{Russell1992} and E0102 O-rich ejecta compositions based on \citet{Blair2000} for the emitting gas in each model. The model parameters are summarized in Table~\ref{tab:models}.

\begin{deluxetable*}{lll}
\tablecaption{Summary of \textsc{Mappings} model grids explored in this work. \label{tab:models}}
\tablewidth{700pt}
\tablehead{
\colhead{Model} & \colhead{Ionizing source / driver} & \colhead{Parameters} 
} 
\startdata
Photoionization & Blackbody (CCO)
& $\log T_{\rm BB}=6.2$–6.5~K, $n_{\rm H}=10$–$10^3$~cm$^{-3}$\\
Photoionization & Thermal bremsstrahlung (hot ejecta) &
$kT \sim 0.79$–0.91~keV, $n_{\rm H}=10$–$10^3$~cm$^{-3}$ \\
Shock & Shock + precursor & $v_s=20$–200 km s$^{-1}$, $n_{\rm H}=10$–$10^3$~cm$^{-3}$ \\
\enddata
\end{deluxetable*}

\begin{figure*}
\plotone{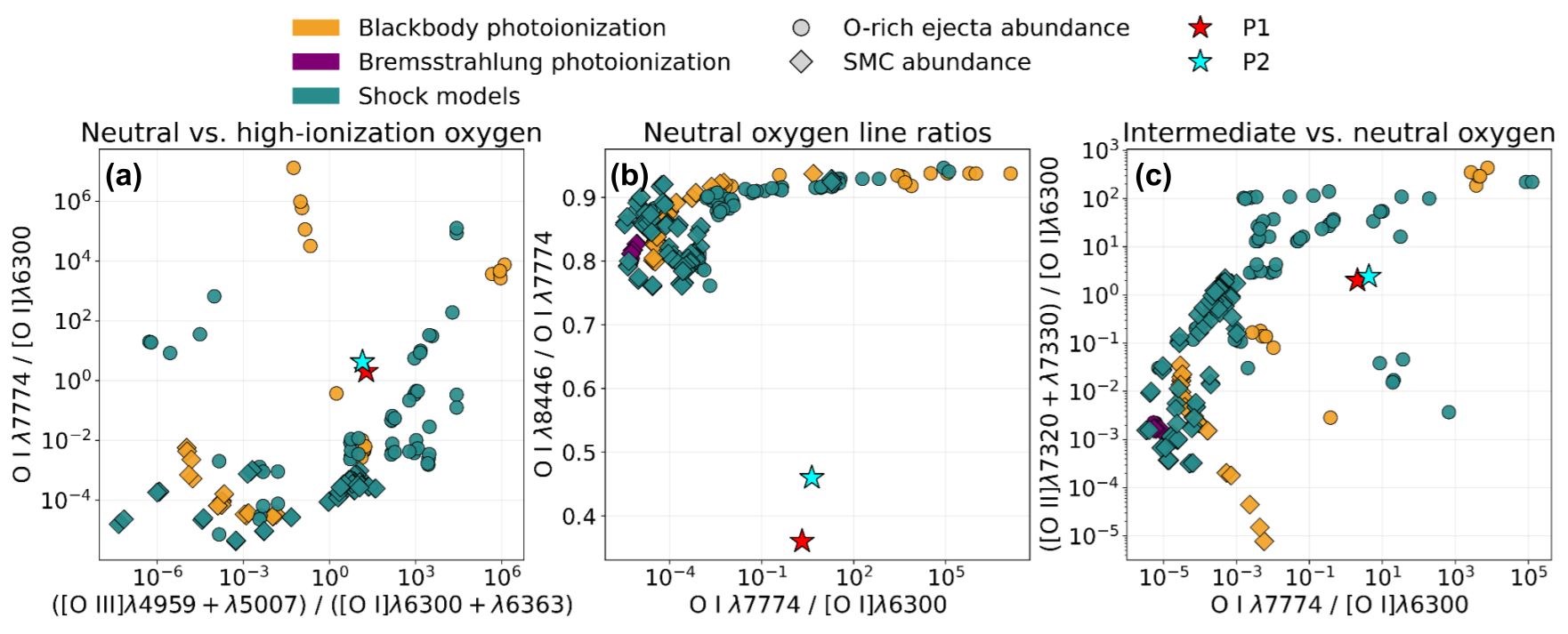}
\caption{Colors distinguish the model type: blackbody photoionization (gold), bremsstrahlung photoionization (purple), and shock models (teal). Marker shapes indicate the assumed abundances of the emitting gas: E0102 O-rich ejecta adopted from \citet{Blair2000} (circles) and SMC composition (diamonds). The observed values for P1 and P2 are shown as stars.
(a) High-ionization versus neutral oxygen diagnostic, showing \ion{O}{1} $\lambda$7774/[\ion{O}{1}]$\lambda$6300 as a function of ([\ion{O}{3}]$\lambda\lambda$4959,5007)/([\ion{O}{1}]$\lambda\lambda$6300,6363).  
(b) Neutral oxygen line ratios, showing \ion{O}{1} $\lambda$8446/\ion{O}{1} $\lambda$7774 versus \ion{O}{1} $\lambda$7774/[\ion{O}{1}]$\lambda$6300.  
(c) Intermediate- versus neutral-ionization diagnostic, showing ([\ion{O}{2}]$\lambda\lambda$7320,7330)/[\ion{O}{1}]$\lambda$6300 as a function of \ion{O}{1} $\lambda$7774/[\ion{O}{1}]$\lambda$6300.}
\label{fig:mappings}
\end{figure*}

Figure~\ref{fig:mappings} shows that no single model reproduces all of the observed diagnostics. The data simultaneously exhibit strong high-ionization emission, traced by large ([\ion{O}{3}]$\lambda\lambda$4959,5007)/([\ion{O}{1}]$\lambda\lambda$6300,6363), and strong neutral oxygen emission, particularly the elevated \ion{O}{1} $\lambda$7774/[\ion{O}{1}]$\lambda$6300 ratio.

Photoionization by a blackbody source with a CCO-like temperature can reproduce the high-ionization ratios, but fails to generate sufficient neutral oxygen emission. In contrast, thermal bremsstrahlung models produce negligible [\ion{O}{3}] emission and therefore do not populate the high-ionization regime (Figure~\ref{fig:mappings}~(a)). Shock models produce a wide range of ionization conditions and can approach either regime, but do not simultaneously reproduce both the high-ionization and neutral diagnostics within the explored parameter space.

The \ion{O}{1} $\lambda$8446/$\lambda$7774 ratio clusters around similar values across all models, whereas the observed values are lower. This discrepancy may be related to the larger intrinsic line widths of $\lambda$7774 relative to $\lambda$8446 in the integrated spectra (see Table~\ref{tab:line_measurements}), which could reflect unresolved kinematic structure and line-of-sight blending.

These results suggest that the emission from P1 and P2 is unlikely to arise from a single homogeneous photoionized or shock-excited component. Instead, the data are consistent with a multi-phase medium in which highly ionized gas coexists with partially neutral, recombining material.

No optical \ion{Ne}{1} lines are present in the \textsc{Mappings} outputs despite their clear detection in the data. This absence reflects limitations in the available atomic data rather than negligible emissivity. Optical \ion{Ne}{1} emission is expected to arise primarily from recombination of \ion{Ne}{2}, but reliable recombination coefficients for these transitions are not implemented in the current \textsc{Mappings} version. As a result, \ion{Ne}{1} is not used to constrain the models, and its detection highlights the need for improved atomic data for low-ionization neon.

%%%%%%%%%%%%%%%%%%%%%%%%%%%%%%%%%%%%%%%%%%%%%%%%%%%%%%%%%%%%%%%%%%%%%%%%%%%%%%%%%%
\section{Discussion} 
\label{sec:discussion}

%%%%%%%%%%%%%%%%%%%%%%%%%%%%%%%%%%%%%%%%%
\subsection{Excitation of the Permitted \ion{O}{1} and \ion{Ne}{1} Emission} \label{subsec:discussion_excitation}

While permitted \ion{O}{1} emission has been observed in SNRs (first reported in Puppis~A; \citealt{Winkler1985}), the detection of optical \ion{Ne}{1} lines in E0102 is unusual and, to our knowledge, has not been reported in other remnants, despite the ubiquity of neon in core-collapse ejecta. This makes \ion{Ne}{1} a particularly strong constraint on the excitation conditions of the optical cavity structure.

The combined detection of permitted \ion{O}{1} and \ion{Ne}{1} emission therefore places stringent constraints on the physical conditions of the emitting region. Any viable excitation mechanism must simultaneously account for:
\begin{enumerate}[nolistsep,topsep=3pt]
    \renewcommand{\labelenumi}{(\roman{enumi})}
    \item the spatial and kinematic correspondence between \ion{O}{1} and \ion{Ne}{1};
    \item the observed \ion{O}{1} $\lambda8446/\lambda7774$ and the detection of \ion{O}{1} $\lambda\lambda$9263,9266;
    \item the coexistence of \ion{O}{1}, \ion{Ne}{1}, [\ion{O}{1}], [\ion{O}{2}], and [\ion{O}{3}]; and
    \item the large intrinsic line widths.
\end{enumerate}

In neutral oxygen, resonant fluorescence can enhance the \ion{O}{1} $\lambda8446$ line via Ly$\beta$ pumping \citep{Bowen1947}. In this process, the wavelength coincidence between Ly$\beta$ and an \ion{O}{1} resonance transition drives a cascade that preferentially strengthens $\lambda8446$. As a result, fluorescence-dominated conditions typically yield $\lambda8446/\lambda7774 \gtrsim 1$ and the ratio can become much larger as pumping strengthens \citep{KB95}.

In the absence of strong fluorescence, the relative strengths of the permitted lines are instead set by a combination of collisional excitation and recombination cascades, which generally produce lower $\lambda8446/\lambda7774$ ratios \citep{BK95}. The observed values across the structure from MUSE NFM data (Figure~\ref{fig:linerat_map}~(a); $\sim$0.2--0.5) are therefore more consistent with collisional and/or recombination-dominated conditions, rather than a fluorescence-dominated regime.

While these ratios point toward collisional and/or recombination-driven excitation of \ion{O}{1}, they do not uniquely constrain the ionization balance of the gas. The detection of \ion{Ne}{1} provides an additional and more stringent constraint on this balance.

With an ionization potential of 21.6~eV, neutral neon is readily ionized in environments exposed to EUV or soft X-ray radiation, such that neon is expected to reside predominantly in higher ionization states (e.g., \ion{Ne}{3}) in strongly photoionized gas. At the same time, producing observable \ion{Ne}{1} emission through recombination requires a substantial population of \ion{Ne}{2}, since the lines arise from recombination cascades. These requirements are not easily satisfied simultaneously: conditions that maintain a significant \ion{Ne}{2} population also tend to suppress the neutral fraction, while environments in which \ion{Ne}{1} can survive may not sustain sufficient recombination emission. This makes it difficult for a purely photoionized or purely recombination-dominated scenario to account for the observed \ion{Ne}{1} emission.

A more plausible interpretation is that the emission arises in partially ionized, non-equilibrium conditions in which \ion{Ne}{1} and \ion{Ne}{2} coexist. In such environments, collisional excitation can efficiently populate the upper levels of both \ion{O}{1} and \ion{Ne}{1}, while recombination contributes by populating excited states, followed by radiative cascades. This interpretation is consistent with the coexistence of species spanning a wide range of ionization states, from neutral \ion{O}{1} and \ion{Ne}{1} to highly ionized [\ion{O}{3}]. In a simple equilibrium nebula, these species would be expected to occupy spatially distinct zones. Their observed spatial and kinematic overlap instead suggests a multiphase medium in which the ionization structure is governed by the dynamical and thermal history of the gas rather than local equilibrium conditions \citep[e.g.,][]{Vink2012}.

Such conditions can naturally arise in environments where the gas is rapidly evolving, including cooling post-shock regions, partially ionized ejecta clumps, and mixing layers within reverse-shocked ejecta. In these settings, variations in density, temperature, and ionization state can coexist on small spatial scales, allowing multiple excitation mechanisms to operate simultaneously.

A limitation of the current modeling is that it treats the emitting gas as a single, homogeneous component in ionization and thermal equilibrium, an assumption that is inconsistent with the multiphase, non-equilibrium conditions implied by the data. This likely explains why no single photoionization or shock model reproduces all of the observed diagnostics. A more complete test of this interpretation would require multi-component modeling that captures the coexistence of gas at different densities, temperatures, and ionization states.

We have subsequently explored a linear combination of photoionization and shock models as a test for multi-component models. While such combinations modestly improve some of the line ratios, they failed to reproduce the full set of observed oxygen diagnostics simultaneously. More importantly, the strongest observational constraint in our dataset is the detection of optical \ion{Ne}{1} emission, which is not currently included in the available \textsc{Mappings} atomic dataset. Any physically meaningful multi-component models would require the implementation and modeling of the relevant Ne I transitions in addition to the oxygen diagnostics. Developing such models is beyond the scope of the present work.

%%%%%%%%%%%%%%%%%%%%%%%%%%%%%%%%%%%%%%%%%
\subsection{Physical Origin of the Cavity} 
\label{subsec:discussion_scenarios}

In this section, we first describe the physical conditions of the emitting gas, and then consider possible mechanisms capable of producing and maintaining the observed cavity structure.

\subsubsection{The emitting gas: a multiphase, partially ionized ejecta structure} 

The observations could be explained as a structure that traces a region of warm, partially ionized ejecta characterized by non-equilibrium conditions. In this scenario, the emission arises from material in which neutral atoms coexist with free electrons, allowing both recombination and collisional excitation processes to contribute.

As shown in Section~\ref{subsec:photo_shock_model}, no single photoionization or shock model reproduces all of the observed diagnostics simultaneously. Models that match the high-ionization ratios fail to generate sufficient neutral oxygen emission, while those that enhance the neutral component do not reproduce the ionized lines. This tension indicates that the emitting region is most likely multiphase.
Cooling post-shock gas naturally produces such conditions, where different ionization states coexist on small scales. The coexistence of \ion{O}{1}, \ion{Ne}{1}, [\ion{O}{1}], [\ion{O}{2}], and [\ion{O}{3}] within the same structure is therefore consistent with a stratified, non-equilibrium medium. The detection of \ion{Ne}{1} is particularly constraining, as it requires partially ionized conditions that are not easily produced in either purely photoionized or equilibrium shock models.

Broad intrinsic line widths further indicate dynamically active gas, consistent with turbulence, unresolved velocity gradients, or the superposition of multiple clumps within the ejecta. These conditions are expected in cooling zones behind shocks, as well as in mixing layers or internally driven shocks, even if the global reverse shock has not yet reached the torus region as argued in \citet{Vogt2018}.

The sharp edge we see is consistent with a shock-driven interface that marks transitions in density and temperature. The cavity morphology described in this work supports a point-like X-ray source driving its formation mechanism. Next, we consider several possibilities for such a source.

\subsubsection{I. A cavity shaped by the CCO candidate}

The optical structure was originally interpreted by \citet{Vogt2018} as being associated with a central X-ray point source, proposed as a CCO candidate. In that scenario, the CCO has not yet been overtaken by the reverse shock, and EUV or soft X-ray radiation from the CCO photoionizes the surrounding material, producing the observed optical structure.

However, the MUSE NFM-AO data and the comparison with \textsc{Mappings} models (Section~\ref{subsec:photo_shock_model}) do not support a purely photoionized origin. Blackbody photoionization models with temperatures consistent with the inferred properties of the CCO cannot simultaneously reproduce the observed combination of strong high-ionization emission and prominent neutral oxygen lines. 

The detection of \ion{Ne}{1} further challenges a purely photoionized origin, as neutral neon is difficult to maintain in a strongly ionized environment. These results indicate that while a central source may contribute locally to the ionization structure, photoionization alone cannot account for the observed combination of neutral and ionized emission. 

The photoionization models discussed above test the radiative impact of the putative CCO. An additional question is whether the CCO could instead shape the cavity through mechanical feedback. If the X-ray source is indeed a neutron star, one possibility is that it drives a relativistic pulsar wind that shapes the surrounding material. The characteristic radius of the cavity inferred from the radial profiles is $\sim$1.6\arcsec, corresponding to $\sim$0.48~pc at the distance of the SMC. This value is substantially larger than the termination shock radii observed in  pulsar wind nebulae, which are $\lesssim$ 0.1 pc even for the most energetic pulsars.

If we nevertheless assume that the cavity is associated with a wind structure, the spin-down power required to balance the surrounding gas pressure is given by $P \approx \dot{E}/(4\pi R^2 c)$; where $\dot{E}$ is the spin-down energy of a putative neutron star,  $R$ is the termination shock radius, and $c$ is the speed of light. For a warm optical gas with temperature of $\sim$10$^4$~K and density of $\sim$10$^2$--10$^3$~cm$^{-3}$ and an assumed termination shock at $\sim$0.48~pc from the X-ray source, the inferred spin-down power is $\dot{E} \sim 10^{37}-10^{38}$~erg\,s$^{-1}$. Such values are several orders of magnitude larger than the known spin-down powers measured for CCOs, which are $\lesssim 10^{33}$~erg\,s$^{-1}$ \citep{2013ApJ...765...58G}. A neutron star with such a high spin-down power would also be expected to produce a prominent pulsar wind nebula and detectable non-thermal emission in radio and/or X-rays, neither of which is observed. Therefore, both the photoionization models and the energetic requirements disfavor a scenario in which the cavity is primarily shaped by a CCO.

Confirming the exact nature of the X-ray source capable of carving such a cavity will require deeper X-ray observations than currently available, in order to lift the existing ambiguities on its spectrum and more cleanly disentangle its signature from that of the surrounding SNR emission.

\subsubsection{II. A cavity related to a binary progenitor or surviving companion}

One way to interpret the cavity is to assume that it marks the effective center of the explosion, in which case its origin could be linked to the progenitor system itself. In this scenario, the cavity would trace material shaped by binary evolution, for example through mass transfer or common-envelope interaction prior to the supernova. Stripped-envelope supernovae are often associated with binary systems, and such interactions can imprint asymmetries or density structures in the surrounding medium.

However, current observational constraints do not support this scenario. \citet{Li2021} found no compelling evidence for a surviving companion at the cavity location, based on searches targeting OB stars and K-type supergiants. Interpreting the cavity as the explosion center of E0102 would therefore require either an undetected or atypical companion.

In addition, the observed physical properties of the emitting gas do not point toward a stellar or circumstellar origin. The emission is dominated by O-rich ejecta, with broad intrinsic line widths and a complex multiphase ionization structure indicative of shock-processed material. These properties differ from those expected for gas shaped primarily by a companion star or pre-explosion mass loss.

While binary interaction may have influenced the global ejecta geometry of E0102, it does not provide a natural explanation for the localized cavity-like feature observed here, particularly given its offset from the independently determined CoE \citep[e.g.,][]{Finkelstein2006,Banovetz2021}.

\subsubsection{III. A cavity produced by an unrelated X-ray source within the remnant}

Given that the cavity is offset from the CoE ($\sim$6.4\arcsec; \citealt{Banovetz2021}), an alternative interpretation is that it is not directly associated with the explosion site or progenitor system. Instead, the structure may be shaped by a separate source embedded within the remnant.

One plausible scenario is that the cavity traces the interaction between the expanding O-rich ejecta and a pre-existing density structure, such as a cavity produced by an unresolved X-ray source within the remnant. The sharp edge morphology can then arise naturally from a density contrast, where the ejecta encounter this source %structure 
and form a shock interface. This interpretation is consistent with the observed properties of the emitting gas, that is dominated by shock-processed ejecta, with broad intrinsic line widths and a multiphase ionization structure. The coexistence of neutral and ionized species further supports a scenario in which the emission arises from dynamically evolving, partially ionized gas.

Similar scenarios have been observed in other remnants, where unrelated objects embedded within the remnant create distinct optical features. For example, in the Vela Juniorct SNR, the CCO nebula has been observed to interact with an unrelated Herbig--Haro system, resulting in a complex structure shaped by this interaction \citep{Suherli2026}.  

In the context of E0102, this scenario provides an explanation for both the offset of the cavity from the center of expansion and its localized morphology. The cavity structure can therefore be understood as a region where the ejecta interact with a pre-existing object, producing the observed sharp boundary and stratified ionization structure.

%%%%%%%%%%%%%%%%%%%%%%%%%%%%%%%%%%%%%%%%%%%%%%%%%%%%%%%%%%%%%%%%%%%%%%%%%%%%%%%%%%
\section{Summary and Future Work} \label{sec:summary}

Using MUSE NFM-AO observations, we re-examine the optical structure in E0102 previously identified as a torus \citep{Vogt2018}, and demonstrate that it is instead a cavity with a sharply defined edge. From the analysis presented in this work, we find that:
\begin{enumerate}
    \item The radial intensity profiles exhibit a sharp rise at a characteristic radius, tracing a well-defined cavity edge. The sharpness of this transition is consistent with a shock-driven interface marking a transition in density and temperature.
    
    \item Spatially resolved line ratios indicate that the emission arises from a structured, multiphase medium. The observed \ion{O}{1} $\lambda8446/\lambda7774$ ratios favor collisional and/or recombination-driven processes, while the detection of \ion{Ne}{1} requires partially ionized conditions in which neutral and ionized species coexist under non-equilibrium conditions.

    \item Comparison with photoionization and shock models shows that no single-component model reproduces the full set of observed diagnostics. This indicates that the emitting gas cannot be described by a simple equilibrium configuration, but instead reflects a combination of physical conditions, likely associated with shock processing in an inhomogeneous medium.
\end{enumerate}

We explored several possible origins for the cavity structure. A purely photoionized cavity driven by the CCO candidate is not supported by the data, as it cannot reproduce the coexistence of neutral and ionized species. Interpreting the cavity as the explosion center associated with a binary progenitor or surviving companion is also disfavored, given the lack of observational evidence for such a companion and the offset from the independently determined center of expansion. Among the source-driven scenarios considered, the most favored is one in which an unrelated X-ray source or pre-existing bubble within the remnant shapes the cavity. Overall, the most consistent interpretation is that the cavity traces the interaction between inhomogeneous O-rich ejecta and a pre-existing unresolved X-ray source within the remnant, producing a localized, shock-defined interface.

Future work will require multi-component, time-dependent models that include non-equilibrium cooling and density inhomogeneities to reproduce the observed coexistence of neutral and ionized species. Additionally, three-dimensional morphological plus kinematic modeling of the cavity and surroundings is needed to unveil the intrinsic geometry of the structure and better probe the relationship between the cavity, X-ray source and surrounding ejecta. Further progress will also depend on improved atomic data for optical \ion{Ne}{1} transitions -- a key finding in our study -- which are not fully treated in current photoionization and shock codes.

Observationally, follow-up studies across multiple wavelengths will provide further constraints. \textit{JWST} spectroscopy will provide access to infrared diagnostics of the ionization and excitation conditions, while deep and sensitive X-ray observations with facilities such as \textit{Chandra} and next-generation missions \citep[e.g. with an \textit{AXIS}-like concept,][]{2023AXIS} will help clarify the nature of the X-ray source, its association with the remnant, and its potential role in shaping the cavity.

%%%%%%%%%%%%%%%%%%%%%%%%%%%%%%%%%%%%%%%%%%%%%%%%%%%%%%%%%%%%%%%%%%%%%%%%%%%%%%%%%%
%% Please use the acknowledgment and contribution environments. This will 
%% be anonomyized when the "anonymous" style option is used. 
\begin{acknowledgments}
This work is based on observations collected at the Europeran Southern Observatory under ESO programme 0104.D--0092(A), P.I.: F.P.A. Vogt. We thank the referee for their constructive feedback that helped improve the clarity of the paper.
The research of JS and SSH is supported by the Natural Sciences and Engineering Research Council of Canada (NSERC) through the Discovery Grants and the Canada Research Chairs program, and by the Canadian Space Agency. CJL is supported by the NSTC grant 114-2112-M-004-001-MY2 from the National Science and Technology Council of Taiwan.
\end{acknowledgments}

%\begin{contribution}
%%This section gives authors the space to recognize author contributions. The text inside this environment is NOT counted towards the total word quanta. At a minimum, manuscripts are expected to include this text:

%All authors contributed ...

%% But authors are expected to provide more specific details, e.g. 
%%
%%SC was responsible for writing and submitting the manuscript.
%%WWM came up with the initial research concept and edited the manuscript.
%%OTS obtained the funding and edited the manuscript.
%%EBF provided the formal analysis and validation. He also edited the manuscript.
%%GEH Supervised the undergraduates, wrote the software and administers the project github and Zenodo repositories.
%%
%% Authors can use the Contributor Role Taxonomy (CRediT) at
%% https://credit.niso.org
%% for ideas on how write a good statement tailored to their needs.

%\end{contribution}

%% To help institutions obtain information on the effectiveness of their 
%% telescopes the AAS Journals has created a group of keywords for telescope 
%% facilities.
%
%% Following the acknowledgments section, use the following syntax and the
%% \facility{} or \facilities{} macros to list the keywords of facilities used 
%% in the research for the paper.  Each keyword is check against the master 
%% list during copy editing.  Individual instruments can be provided in 
%% parentheses, after the keyword, but they are not verified.
\facilities{VLT:Yepun (MUSE)}

%% Similar to \facility{}, there is the optional \software command to allow 
%% authors a place to specify which programs were used during the creation of 
%% the manuscript. Authors should list each code and include either a
%% citation or url to the code inside ()s when available.
\software{
            \textsc{Scipy} \citep{2020SciPy-NMeth},
            \textsc{Astropy} \citep{astropy:2013,astropy:2018,astropy:2022},
            \textsc{brutifus} \citep{brutifus} (a Python module to process datacubes
                from integral field spectrographs, that relies on \textsc{statsmodel} \citep{seabold2010_2}, \textsc{matplotlib}, \textsc{astropy}, and \textsc{photutils}, an affiliated package of \textsc{astropy} for photometry), 
            \textsc{Matplotlib} \citep{Hunter:2007},
            \textsc{CosmosCanvas} \citep{CosmosCanvas}, and
            \textsc{velociwrap} \citep{velociwrap}.
          }

%% Appendix material should be preceded with a single \appendix command.
%% There should be a \section command for each appendix. Mark appendix
%% subsections with the same markup you use in the main body of the paper.
%%
%% Each Appendix (indicated with \section) will be lettered A, B, C, etc.
%% The equation counter will reset when it encounters the \appendix
%% command and will number appendix equations (A1), (A2), etc. The
%% Figure and Table counter will not reset.

\appendix

\section{MUSE NFM-AO Observation Log} \label{app:nfm_obs}

Table~\ref{tab:obslog} summarizes the MUSE NFM-AO observations used in this work, including observing time, observation block (OB) identifiers, field configuration, position angle, exposure time, and observing conditions.

\begin{deluxetable*}{lccccccl}
\tablecaption{MUSE NFM-AO Observation Log \label{tab:obslog}}
\tablewidth{700pt}
\tabletypesize{\scriptsize}
\tablehead{
\colhead{Observing Time} & \colhead{OB ID\tablenotemark{a}} & \colhead{Field} & 
\colhead{PA} & \colhead{Exposure Time} & \colhead{Airmass\tablenotemark{b}} & 
\colhead{DIMM Seeing\tablenotemark{c}} & \colhead{Notes} \\ 
\colhead{(UTC)} & \colhead{} & \colhead{} & 
\colhead{(deg)} & \colhead{(s)} & \colhead{} & 
\colhead{(arcsec)} & \colhead{} 
} 
\startdata
2020-11-12 02:36:51.007 & 2493962 & B  &   135 &    1495 &     1.47 &   0.59 &   \\
2020-12-13 02:19:57.345 & 2493951 & B  &    45 &     231.52 &     1.56 &   0.50 & Excluded from final mosaic\tablenotemark{d}  \\
2021-08-17 07:56:02.757 & 2493968 & B  &   315 &     413.81 &     1.47 &   0.44 & Excluded from final mosaic\tablenotemark{d}  \\
2021-08-17 08:11:54.621 & 2493968 & B  &   315 &    1495 &     1.47 &   0.51 &   \\
2021-08-17 08:39:17.124 & 2493968 & B  &   315 &    1495 &     1.48 &   0.51 &   \\
2021-09-06 05:28:37.454 & 2493971 & C  &    77 &    1495 &     1.51 &   0.62 &   \\
2021-09-21 02:24:30.840 & 2494035 & C  &   167 &     395 &     1.73 &   0.35 &  Excluded from final mosaic\tablenotemark{d} \\
2021-11-03 05:26:23.389 & 2494044 & C  &   347 &    1495 &     1.61 &   0.35 &   \\
2021-11-03 05:52:56.037 & 2494044 & C  &   347 &    1495 &     1.67 &   0.34 &   \\
2022-01-02 01:06:26.719 & 2489756 & A  &    65 &    1495 &     1.57 &   0.58 &   \\
2022-01-02 01:32:59.616 & 2489756 & A  &    65 &    1495 &     1.62 &   0.47 &   \\
2022-08-24 05:09:23.942 & 2493945 & A  &   245 &    1495 &     1.60 &   0.60 &   \\
2022-08-24 05:35:56.641 & 2493945 & A  &   245 &    1495 &     1.56 &   0.59 &   \\
2022-08-25 06:19:15.847 & 2493962 & B  &   135 &    1495 &     1.50 &   0.47 &   \\
2022-08-25 06:45:48.635 & 2493962 & B  &   135 &    1495 &     1.49 &   0.46 &   \\
2022-08-25 07:24:47.612 & 2493965 & B  &   225 &    1495 &     1.47 &   0.53 &   \\
2022-08-25 08:04:14.292 & 2493965 & B  &   225 &    1495 &     1.48 &   0.61 & Excluded from final mosaic\tablenotemark{d}  \\
2022-08-27 05:53:45.337 & 2493971 & C  &    77 &    1495 &     1.52 &   0.38 &   \\
2022-08-27 06:20:18.177 & 2493971 & C &    77 &    1495 &     1.50 &   0.51 &   \\
2022-08-28 04:55:20.661 & 2494035 & C  &   167 &    1495 &     1.60 &   0.44 &   \\
2022-08-28 05:21:54.148 & 2494035 & C  &   167 &    1495 &     1.56 &   0.27 &   \\
2022-08-28 05:56:35.194 & 2494041 & C  &   257 &    1495 &     1.51 &   0.34 & Excluded from final mosaic\tablenotemark{d}  \\
2022-08-28 06:23:08.268 & 2494041 &  C &   257 &    1495 &     1.49 &   0.49 &   \\
2022-08-28 06:54:42.252 & 2493942 & A  &   155 &    1495 &     1.48 &   0.39 &   \\
2022-08-28 07:21:15.211 & 2493942 & A  &   155 &    1495 &     1.47 &   0.33 &   \\
2022-08-29 04:37:23.659 & 2493951 & B  &    45 &    1495 &     1.63 &   0.55 &   \\
2022-08-29 05:03:55.661 & 2493951 & B  &    45 &    1495 &     1.58 &   0.37 &  Excluded from final mosaic\tablenotemark{d} \\
\enddata
\tablenotetext{a}{Observation Block identifier.}
\tablenotetext{b}{The airmass at the start of the observation.}
\tablenotetext{c}{Differential Image Motion Monitor; The observatory seeing at the start of the exposure.}
\tablenotetext{d}{Observations marked as excluded were not included in the final mosaic due to lower data quality (e.g., short exposure time, poorer AO correction, or residual artifacts).}
\end{deluxetable*}

\section{Single-line Channel Maps} 
\label{app:chanmap}

For the construction of channel maps, the continuum-subtracted datacube was interpolated onto a uniform velocity grid with a spacing of $\Delta v_{\text{los}} = 50~\mathrm{km\,s^{-1}}$ using linear interpolation along the spectral axis. The interpolation conserves integrated line fluxes to within numerical precision and does not affect the inferred spatial or kinematic properties discussed in this work. This rebinning was performed solely for visualization and comparative analysis; all quantitative measurements were carried out on the original spectral sampling.

\begin{figure*}[p]
\centering
\makebox[\textwidth][c]{%
  \rotatebox{90}{%
    \begin{minipage}{1.35\textwidth}
      \centering
      \includegraphics[width=\textwidth]{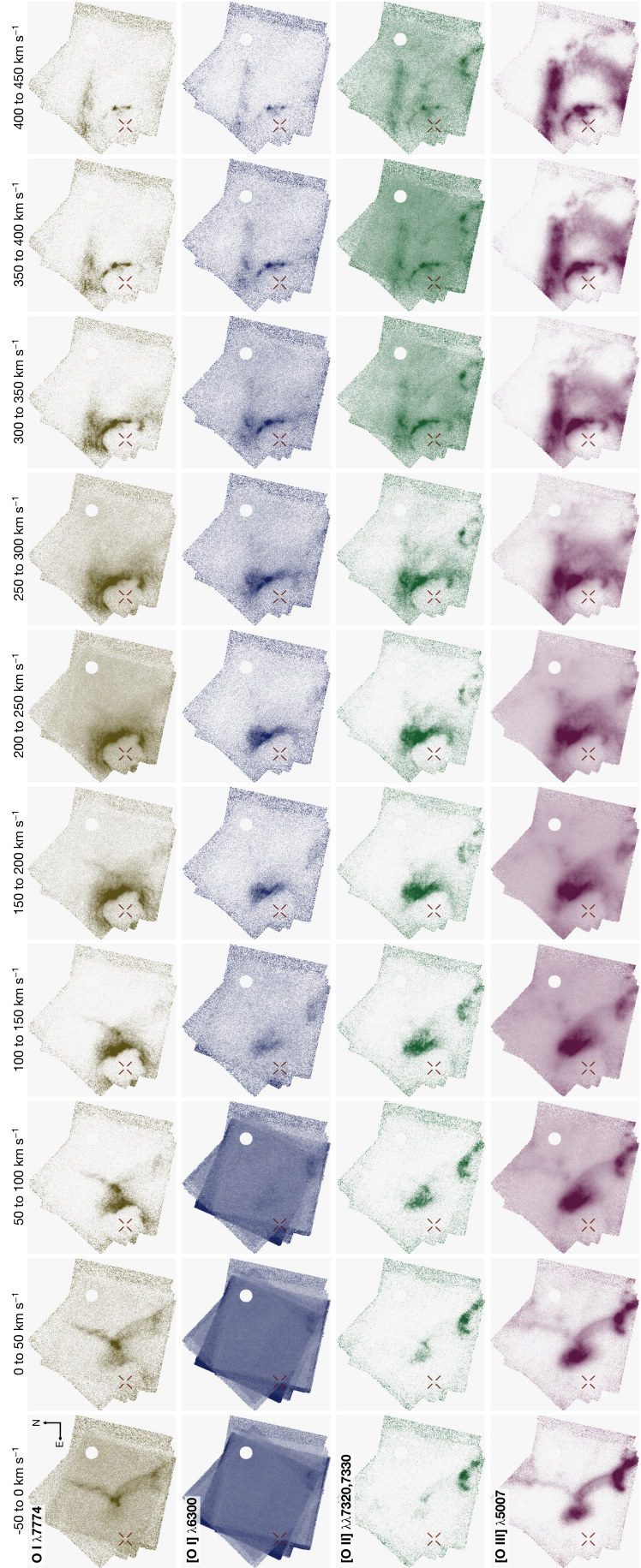}
      \caption{Channel maps of \ion{O}{1} $\lambda$7774 (top row), [\ion{O}{1}]$\lambda$6300 (second row), [\ion{O}{2}]$\lambda$7320,7330 (third row), and [\ion{O}{3}]$\lambda$5007 (bottom row), constructed in 50~km\,s$^{-1}$ bins from $-50$ to $+450~\mathrm{km\,s^{-1}}$ (left to right). The dark-red crosshair marks the position of the X-ray source, and the white circle indicates a masked foreground star. The orientation on the sky is shown in the upper-left panel.}
      \label{fig:chanmap_all}
    \end{minipage}%
  }%
}
\end{figure*}

\section{The Cavity Radius} 
\label{app:radius}

\begin{figure*}
\plotone{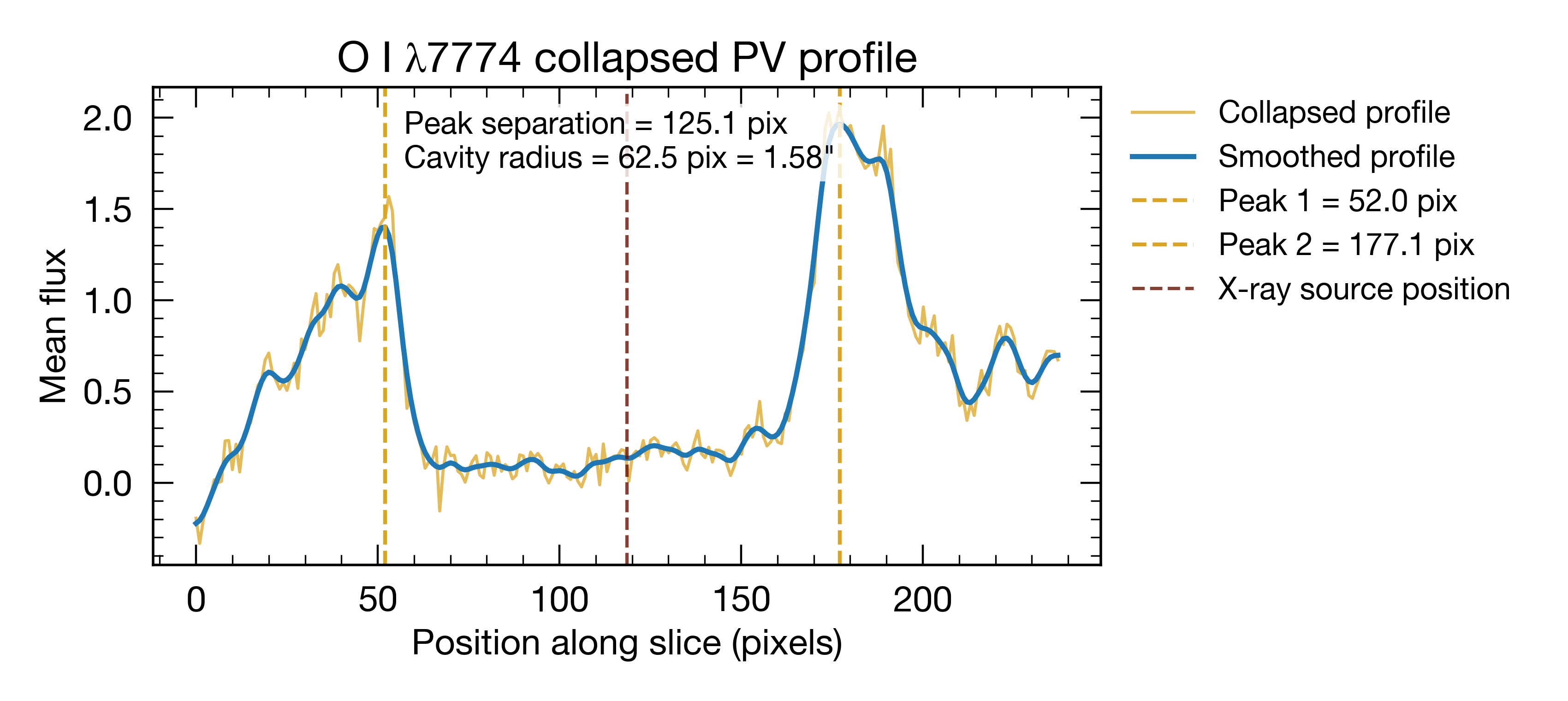}
\caption{Collapsed PV profile of \ion{O}{1} $\lambda7774$ along the slice. The gold curve shows the mean flux profile and the blue curve shows a Gaussian-smoothed version used to identify the dominant peaks. The yellow dashed vertical lines mark the positions of the two peaks and the red dashed line marks the position of the X-ray source along the slice.
}
\label{fig:cavity_radius}
\end{figure*}

To quantify the projected scale of the cavity, we collapse the PV diagrams over the velocity range $-50 \le v_{\mathrm{los}} \le +450~\mathrm{km\,s^{-1}}$ and examine the resulting one-dimensional intensity profile along the slice (Figure~\ref{fig:cavity_radius}).

The collapsed profile (gold curve) traces the mean flux as a function of position, while a Gaussian-smoothed version (blue curve) is used to suppress small-scale fluctuations and more robustly identify the dominant peaks. We identify two prominent peaks corresponding to the projected boundaries of the cavity and measure their separation along the slice. This yields a peak separation of $\sim$125 pixels, implying a characteristic cavity radius of $\sim$62.5 pixels, corresponding to $\sim$1.6\arcsec\ or 0.48~pc at the SMC distance of 62~kpc.

%% For this sample we use BibTeX plus aasjournalv7.bst to generate the
%% the bibliography. The sample7.bib file was populated from ADS. To
%% get the citations to show in the compiled file do the following:
%%
%% pdflatex sample7.tex
%% bibtext sample7
%% pdflatex sample7.tex
%% pdflatex sample7.tex

\bibliography{e0102}{}
\bibliographystyle{aasjournalv7}
%\bibliographystyle{aasjournal}

%% This command is needed to show the entire author+affiliation list when
%% the collaboration and author truncation commands are used.  It has to
%% go at the end of the manuscript.
%\allauthors

%% Include this line if you are using the \added, \replaced, \deleted
%% commands to see a summary list of all changes at the end of the article.
%\listofchanges

\end{document}